\documentclass[preprint]{elsarticle}
\usepackage{footmisc}
\makeatletter
\let\@fnsymbol\@alph
\makeatother
\usepackage[english]{babel}
\usepackage[utf8]{inputenc}
\usepackage{fancyhdr}
\usepackage{amsmath}
\usepackage{amssymb}
\usepackage{upgreek}
\usepackage{graphicx}
\usepackage{endnotes}
\usepackage{float}
\usepackage{caption}
\usepackage{subcaption}
\usepackage[colorinlistoftodos]{todonotes}
\usepackage{url}
\usepackage{lineno}
\usepackage{appendix}
\usepackage[colorlinks]{hyperref}
\AtBeginDocument{%
  \hypersetup{%
    linkcolor=red,%
    citecolor=green,%
    urlcolor=cyan,%
  }%
}%

\usepackage{geometry}
\newcommand{\qgsp}{{\sc qgsp\_bert}}
\newcommand{\ftfp}{{\sc ftfp\_bert}}
\newcommand{\qbbc}{{\sc qbbc}}
\newcommand{\ecal}{Si-W ECAL}
\newcommand{\ecalp}{\ecal\ physics prototype}
\newcommand{\tfa}{track-finding algorithm}
\newcommand{\ep}{$\varepsilon$ parameter}

\newcommand{\geantfour}{{\sc geant4}}

\usepackage{fancyhdr}
\fancypagestyle{pprintTitle}{%
\lhead{} \chead{}\rhead{CALICE-PUB-2019-002}
\lfoot{}\cfoot{}\rfoot{}

}

\begin{document}

\begin{frontmatter}

%\title{ {\LARGE\bf A precise determination of top quark electro-weak couplings at the ILC operating at $\roots=500\,\GeV$}}
%\title{\LARGE\bf Tracks of hadronic showers in the CALICE \ecalp \\ \bigskip {\Large\bf The CALICE Collaboration}}
\title{\LARGE\bf Characterisation of different stages of hadronic showers using the CALICE \ecalp \\ \bigskip {\Large\bf The CALICE Collaboration}}

%\author{{\Large The CALICE Collaboration}\\ \bigskip}

%\address{}
%Authorlist
%\normalfont \bf
\author[Bergen]{ G.\,Eigen}
\author[Birmingham]{T.\,Price}
\author[Birmingham]{N.K.\,Watson}
\author[Birmingham]{A.\,Winter}
\author[Kyungpook]{Y.Do} 
\author[Kyungpook]{A.Khan}
\author[Kyungpook]{D.Kim}
\author[NICADD]{G. C.\,Blazey} 
\author[NICADD]{A.\,Dyshkant} 
\author[NICADD]{K.\,Francis} 
% ...
 
%Institute addresses
\address[Bergen]{{\it University of Bergen, Inst.\, of Physics, Allegaten 55, N-5007 Bergen, Norway}}
\address[Birmingham]{ {\it University of Birmingham, School of Physics and Astronomy, Edgbaston, Birmingham B15 2TT, UK}}
\address[Kyungpook]{{ \it Department of Physics, Kyungpook National University, Daegu, 702-701, Republic of Korea}}  
\address[NICADD]{ {\it NICADD, Northern Illinois University, Department of Physics, DeKalb, IL 60115, USA }}
% ...
\newpage
%Supplementary information
\fntext[oskar]{Now at University of Hawaii at Manoa, High Energy Physics Group, 2505 Correa Road, HI, Honolulu 96822, USA}
%\fntext[danilov]{Also at Moscow Institute of Physics and Technology, 9 Institutskiy per., Dolgoprudny, Moscow Region, 141701, Russian Federation}
\fntext[amjad]{Now at Department of Physics and Astronomy, University College London, Gower Street, London WC1E 6BT, UK }
\fntext[bilokin]{Now at IPHC Strasbourg, 23 rue du loess, BP28, 67037 Strasbourg cedex 2}
\fntext[doublet]{Now at IUT d'Orsay (Universit\'e Paris-Sud), Plateau de Moulon, 91400 Orsay, France}
\fntext[hengneli]{Now at South China Normal University, 55 Zhong Shan Da Dao Xi, Tianhe District, 510631 Guangzhou, Guangdong, China}
\fntext[cornat]{Now at Laboratoire de Physique Nucl\'eaire et de Hautes Energies (LPNHE), UPMC, UPD, CNRS/IN2P3, 4 Place Jussieu, 75005 Paris, France }

%Corresponding author(s)
\cortext[cor1]{Corresponding author: sviatoslav.bilokin@iphc.cnrs.fr}
\cortext[cor2]{Corresponding author: poeschl@lal.in2p3.fr}

\begin{abstract}
A detailed investigation of hadronic interactions is performed using $\uppi^-$-mesons with energies in the range 2--10 GeV incident on a high granularity silicon-tungsten electromagnetic calorimeter. The data  were recorded at FNAL in 2008. The region in which the $\uppi^-$-mesons interact with the detector material and the produced secondary particles are characterised using a novel track-finding algorithm that reconstructs tracks within hadronic showers in a calorimeter in the absence of a magnetic field. The principle of carrying out detector monitoring and calibration using secondary tracks is also demonstrated.

%This article presents a detailed investigation of  hadronic interactions in a highly granular electromagnetic calorimeter prototype
%by means of a novel \tfa. The analysis studies the characteristics of the interaction region and of the tracks that emerge from the interaction of a primary hadron with the detector material. 
%This note presents preliminary results obtained analyzing the \ecalp\ data collected during the test-beam running periods in 2008 year at FNAL.
%Data taken at FNAL in 2008 are compared with predictions of three \geantfour\ simulation models.
%Results of track-finding algorithm on the experimental data was compared to {\sc Geant}4 simulations with different physics lists. 
%Several new observables were used to show that 
%The simulation models provide a good description of the data in terms of new observables, available through the detailed analysis of the secondary particles; the predictions are within 15\% of the data, and for many observables much closer. A consistency check by selecting clean MIP like tracks demonstrates that the current algorithm allows, in principle, for detector calibration and monitoring.  
%Main systematic effects were studied using electron and muon samples. 
%This work is based on the study published in \cite{bib:Naomi}.
\end{abstract}

\end{frontmatter}

%\begin{titlepage}
%\begin{flushright}
%Draft CALICE Paper 028\\
%          Version 2.4\\
%           \today\\
%\end{flushright}
%\maketitle 
%\bigskip\bigskip\bigskip\bigskip\bigskip\bigskip

%\begin{center}
%\huge \bf 
%Tracks of hadronic showers in the \ecalp

%\end{center}\bigskip\bigskip 
%\begin{center}{
%{\LARGE The CALICE Collaboration}

%\footnote{Corresponding authors: \\ Roman P\"oschl: {\tt poeschl@lal.in2p3.fr}, Sviatoslav Bilokin: {\tt bilokin@lal.in2p3.fr}}}
%\end{center}\bigskip\bigskip
%\bigskip%\begin{center}{\large  Abstract}\end{center}

%\begin{bottompar}
%\begin{center}
%\vspace{2cm}
%{\sl This note contains preliminary CALICE results, and is for the use
%of members of the CALICE Collaboration and others to whom permission
%has been given.}
%\end{center}
%\end{bottompar}
%\end{titlepage}

%\thispagestyle(fancy)
%\newpage
\tableofcontents

%------------------Introduction---------------------%
\section{Introduction}
%General introduction

The design of particle detectors at future high-energy physics experiments and, in particular, at linear colliders is oriented towards the usage of Particle Flow Algorithms (PFA) for the event reconstruction. 
These algorithms aim to achieve good jet energy resolution by reconstructing individual particles and hence require high granularity calorimeters~\cite{Brient:2002gh, Morgunov:2002pe, Thomson:2009rp}. 

The primary objective of the CALICE (Calorimeter for the Linear Collider Experiment) collaboration is the development, construction and testing of highly granular hadronic and electromagnetic  calorimeters for future particle physics experiments.

A detailed study of the calorimeter response to particle interactions is necessary to verify existing Monte Carlo simulation models and to build reliable PFA. 
This implies the precise simulation and reconstruction of the interaction of neutral and charged hadrons using the subsequent particle cascade.

This article presents a detailed study of ${\uppi^-}$-meson interactions in the CALICE Silicon-Tungsten Electromagnetic Calorimeter (\ecal) physics prototype \cite{Anduze:2008hq}.
The \ecal\ was tested at Fermi National Accelerator Laboratory (FNAL) in 2008 using a beam of $\uppi^-$-mesons in the energy range from 2 to 10\,GeV. 
%High energy jets are predominantly composed of neutral and charged pions within this energy range and therefore the performance of Monte Carlo simulations with these particles is important.
The highly granular structure of the \ecal\ enables both a detailed measurement of hadronic showers in terms of integral observables \cite{Adloff:2010xj, Bilki:2014uep} as well as deeper studies of the interactions between hadrons and the absorber material, such as the characterisation of the interaction region and the analysis of secondaries emerging from the interaction. The tracks produced by these secondaries are reconstructed using a new simple \tfa . The resulting observables are used to compare data with predictions from several \geantfour\ simulation models~\cite{Allison:2006ve, Allison:2016lfl}. The analysis complements studies presented in~\cite{Adloff:2013vra} and~\cite{TheCALICE:2017ubb} for tracking in CALICE prototypes of hadronic calorimeters.  

%and differential observables, like kinematics of secondary particles, that emerge from hadron-detector interactions.

%TODO: insert new results of study if any
%This study is done as an extension of Ref. \cite{bib:Naomi}, where Monte Carlo simulation are compared with data using global observables like lateral and longitudinal extension of the hadronic cascade immediately after the first interaction.
%and they can create energy depositions outside a main hadronic showers. 
%The reconstruction of these secondary tracks provides a new set of observables that can:
%\begin{itemize}
%\item extend Monte-Carlo comparison with experimental data by using kinematic observables of secondary particles;
%\item improve reconstruction of the energy of neutral hadrons by detecting distant energy depositions using track information inside calorimeter;
%\item improve Particle ID algorithms by introducing more detailed shower shape information for identification of hadron interaction in the \ecal.
%\end{itemize}
%||||||||||||||||||||||||ECAL|||||||||||||||||||||||%

\section{The \ecalp}\label{sec:ecalp}
\label{sec:ecal}
\begin{figure}[H]
\centering
\includegraphics[width=0.5\textwidth]{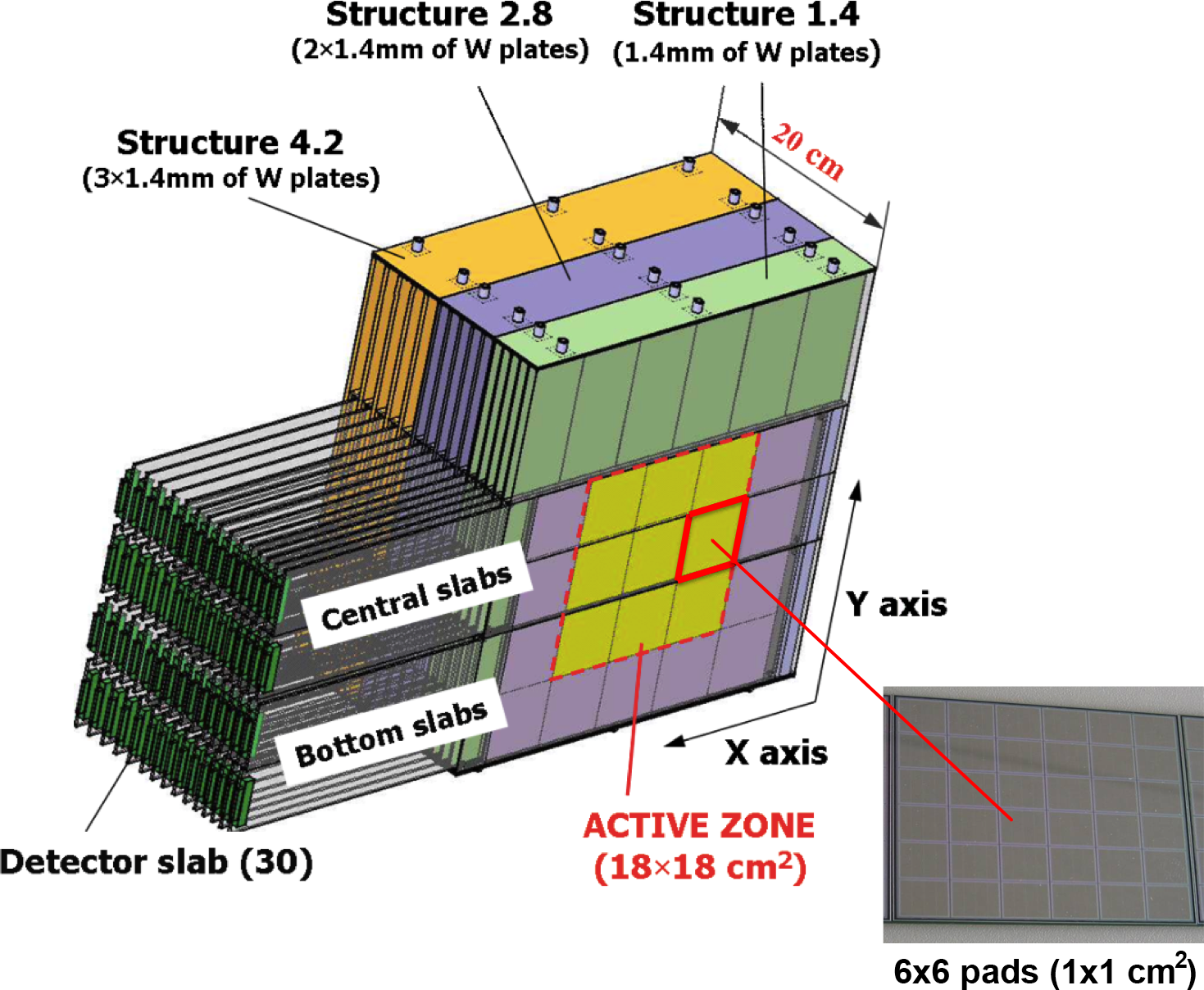}
\caption{\label{fig:ECAL-scheme} \sl A schematic view of the \ecalp.}
\end{figure}

The \ecalp\ has a sandwich-like structure comprising 30 layers of silicon as the active material, alternating with tungsten as the absorber material. The active layers are made of Si wafers segmented in 1 $\times$ 1\,cm$^2$ pads. As shown in Fig. \ref{fig:ECAL-scheme}, each wafer consists of a square of 6 $\times$ 6 pads and each layer is a matrix of 3 $\times$ 3 of these wafers resulting in an active zone of 18 $\times$ 18\,cm$^2$.
The \ecal\ is subdivided into three modules of ten layers. The tungsten depth per layer is 1.4\,mm (0.4 radiation lengths, $X_0$) in the first module, 2.8\,mm in the second and 4.2\,mm in the third. The total thickness corresponds to 24 $X_0$ and about one nuclear interaction length $\lambda_I$. Therefore more than half of the hadrons are expected have a primary interaction within the detector volume.
%This frame is used for the clustering and tracking algorithm, described in Sec. \ref{sec:track}.
%The direction of $z$-axis is parallel to the primary particle momentum.
A more detailed description of the prototype can be found in Ref.~\cite{Anduze:2008hq}.

For the analysis presented in this article it is convenient to introduce a unit grid based on the \ecal\ pad identifiers according to 
%pad identifier
\begin{equation}
	\vec{x} = (x,y,z)=\left
\{
\begin{array}{c}
x= 0 .. 17 \\ 
y=0 .. 17 \\
z = 0 .. 29,
\end{array}
\right.
\end{equation}
where pad counting starts in the bottom right pad, see Fig.~\ref{fig:ECAL-scheme}. Distances in this grid are measured in {\em grid units}, ${\rm g.u.}$

%|||||||||||||||||||Description of a dataset||||||||||||||||||||%

\section{Data and Monte Carlo samples}
\label{sec:data}
\subsection{Experimental setup at FNAL}\label{sec:fnal}
\begin{figure}[H]
\centering
\includegraphics[width=0.95\textwidth]{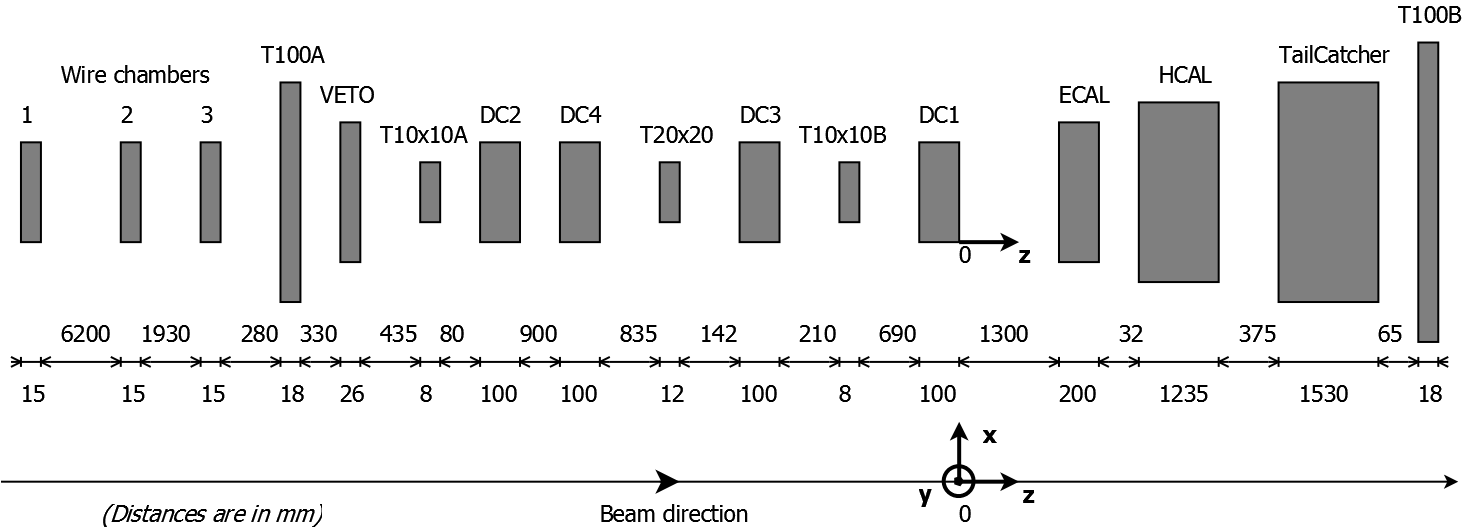}
\caption{\label{fig:fnal-beamline} \sl Plan view of the beam line at FNAL. Distances (not to scale) are in mm.}
\end{figure}

The test beam measurements were carried out at the Fermilab Test Beam Facility\footnote{\label{note1}Fermilab Test Beam Facility web page: \url{http://www-ppd.fnal.gov/MTBF-w}}, FTBF, at FNAL in May and July 2008. A schematic overview of the beam line is given in Fig.~\ref{fig:fnal-beamline}. The \ecal\ was placed in front of two other CALICE physics prototypes, the analogue hadronic calorimeter (HCAL) \cite{collaboration:2010hb} and a TailCatcher~\cite{CALICE:2012aa}. In both steel is used as absorber. Sensors are scintillator tiles (HCAL) or scintillator strips read out by silicon-photomultipliers. The beam-line also included wire chambers (WC1-3), drift chambers (DC1-4) and scintillator counters of different sizes, named T100(A,B), VETO, T20x20 and T10x10(A,B). The latter two cover an area of 10 $\times$ 10 cm$^2$ each and are used for triggering on beam particles analysed for this article. Finally, two Cherenkov detectors for particle identification are located upstream of the Wire Chamber 1. 

The chosen coordinate system is right-handed with the $z$-axis pointing along the beam direction and the $y$-axis being vertical.
The data analysed in this article comprise runs with $\uppi^-$-mesons with energies of 2, 4, 6, 8 and 10 GeV.

\subsection{Monte Carlo simulations \label{mc-models} }

%Due to the complicated nature of hadronic interactions a precise description of hadronic showers in simulations is difficult to achieve. 
Monte Carlo simulations were carried out within the Mokka framework~\cite{MoradeFreitas:2002kj}, which provides the geometry interface to \geantfour. 
There are several models of hadronic interactions available within \geantfour\ that are combined into simulation models.
Each hadronic interaction model has its own theoretical basis valid mainly in a specific energy range of hadrons. In this analysis, three simulation models contained in \geantfour\ Version 10.1 are compared with the data:
\begin{itemize}
\item \ftfp\ uses the Bertini Cascade Model~\cite{Wright:2015xia} and the Fritiof String Model~\cite{Andersson:1986gw, NilssonAlmqvist:1986rx} where the first is used for hadron energies below and the second for hadron energies above 4.5\,GeV;

\item \qgsp\ uses the Bertini Cascade Model at energies below,  and the Fritiof String Model for energies above 9.9 GeV. The Fritiof String Model replaces the LHEP parametrisation that was employed until Version 9.6 of \geantfour;

\item \qbbc\ uses also the Bertini Cascade Model for energies below and the Fritiof String Model for energies above 9.9 GeV but interpolates in a larger transition region (for protons and neutrons below 1.5\,GeV the Binary Cascade Model~\cite{Folger:865824} is used).
\end{itemize}

\begin{figure}[H]
\centering
\includegraphics[width=0.95\textwidth]{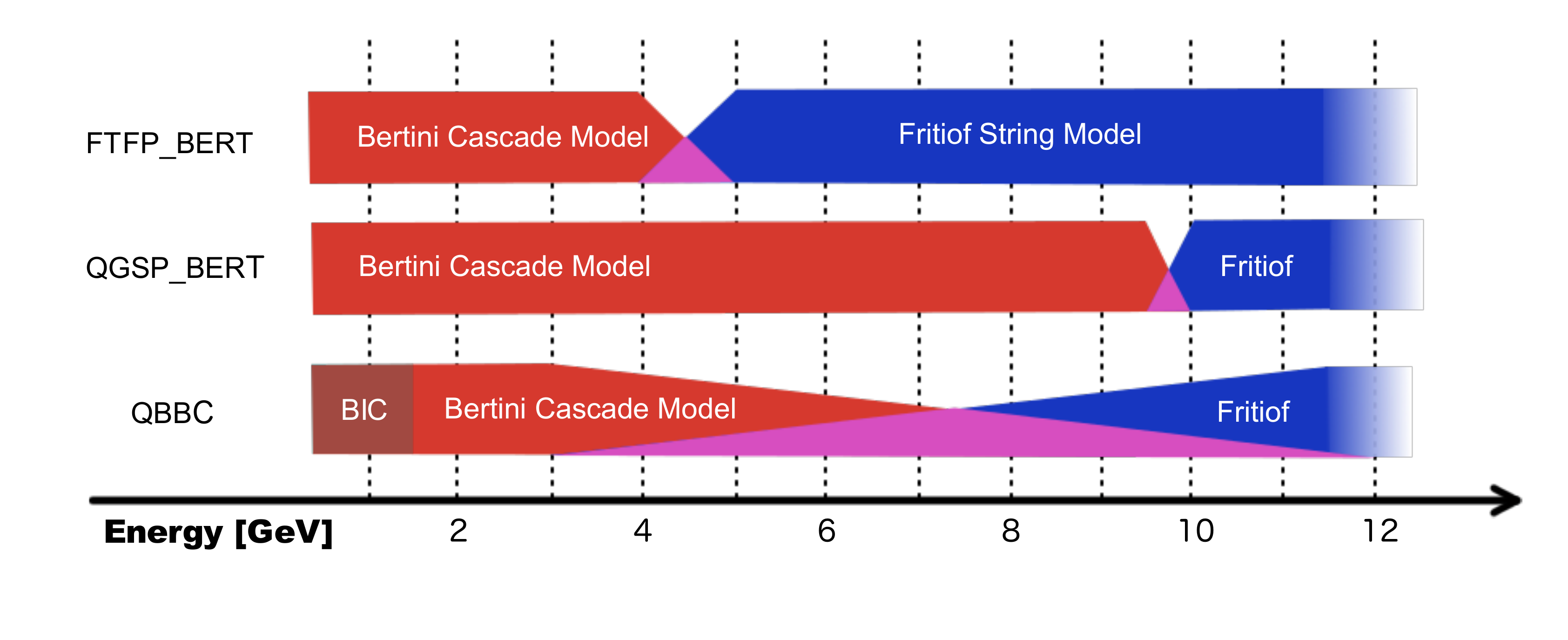}
\caption{\label{fig:g4-val101} \sl Illustration of the validity ranges of the three tested \geantfour\ hadronic interaction models as contained in \geantfour\ Version 10.1.}
\end{figure}

The validity ranges of hadronic interaction models in the three simulation models are illustrated in Fig.~\ref{fig:g4-val101}. More information about these and other simulation models can be found in Ref.~\cite{bib:G4pl}.

\subsection{Event selection and preprocessing}

The FNAL $\uppi^-$ test beam is contaminated with $\upmu^-$ and $\mathrm{e^-}$, in particular at lower energies. 
%where the beam is dominated by $\mathrm{e^-}$. 
At 2 GeV the beam contains about 5\% $\uppi^-$-mesons and 70\% electrons. 
Events are triggered using the signals from the two scintillator counters T10x10A and T10x10B upstream of the \ecal\ and $\uppi^-$-mesons are identified by using Cherenkov counters.
The response of the \ecal\ to charged particles was calibrated with a 32\,GeV $\upmu^-$ beam \cite{li:tel-00430432} and the deposited energy is converted into units of most probable energy depositions, called MIP hereafter, by particles with behaviour that is approximately minimally ionising. The deposited energy measured in a pad is called hit.
%Muons penetrate the whole detector volume with a nearly identical energy loss rate which is minimal for the beam energy used. 
%These $\upmu^-$ are so-called minimum ionising particles or MIP and their mean energy loss in the active medium of a pad defines the energy unit MIP.

To select $\uppi^-$-meson showers, data and simulation samples are required to satisfy similar criteria to those of Refs.~\cite{Bilki:2014uep, doublet:tel-00657967}, as below:  
\begin{itemize}
\item Selection criteria are applied to reject multi-particle events caused by beam impurities or products of decays or upstream interactions of beam particles;
\item A lower threshold of 0.6 MIP is chosen to remove noise hits in the \ecal;
\item A hit is classified as being isolated if all the 26 pads in the surrounding cube (in g.u.) have no signal above the noise threshold. The analysis presented in this article uses the non-isolated hits that remain after this removal. The term `hits` will continue to be used to indicate only non-isolated hits in the following.   
%\item Hits are further rejected if  if all the 26 pads in the surrounding cube (in g.u.) have no signal above the noise threshold.
\item A total of at least 25 hits in the \ecal\ is required to remove particles with large incident angle;
\item For the event selection the hit coordinates $x_{hit}$ and $y_{hit}$ are defined in the coordinate frame according to Fig.~\ref{fig:fnal-beamline}. The barycentres of the transverse coordinates $\bar{x}_{hit}$ and $\bar{y}_{hit}$ of the hits are calculated as:
\begin{eqnarray}
\label{eq:barycentre}
	\bar{x}_{\mathrm{hit}} = \frac{\displaystyle \sum_{\mathrm{hits}} x_{\mathrm{hit}}\,E_{\mathrm{hit}}}{\displaystyle \sum_{\mathrm{hits}} E_{\mathrm{hit}}} 
    \text{ and }
   	\bar{y}_{\mathrm{hit}} = \frac{\displaystyle \sum_{\mathrm{hits}} y_{\mathrm{hit}}\,E_{\mathrm{hit}}}{\displaystyle \sum_{\mathrm{hits}} E_{\mathrm{hit}}},
\end{eqnarray}
where $E_{\mathrm{hit}}$ is the energy of a hit in MIP units, and the sums run over all hits in the calorimeter.
The event is accepted if $-50\,{\mathrm {mm}} < \bar{x}_{\mathrm {hit}} < 50\,{\mathrm {mm}}$ and $-50\,{\mathrm {mm}} < \bar{y}_{\mathrm {hit}} < 50\,{\mathrm {mm}}$ to
reduce lateral shower leakage;
\item  Initially, the interaction layer $i$ is identified as the first of three consecutive layers for which
\begin{equation}
	E_i > E_{\mathrm {cut}}, E_{i+1} > E_{\mathrm {cut}}  \text{ and } E_{i+2} > E_{\mathrm{cut}}.
\end{equation}
with $E_i$ being the total energy of layer $i$;

This simple condition is inefficient at low energies and is extended by the following relative energy increase
\begin{equation}
	\frac{E_i+E_{i+1}}{E_{i-1} + E_{i-2}} > F_{\mathrm{cut}} \text{ and } 
    \frac{E_{i+1}+E_{i+2}}{E_{i-1} + E_{i-2}} > F_{\mathrm{cut}}. 
\end{equation}
The variables $E_{\mathrm{cut}}$ and $F_{\mathrm{cut}}$ are free parameters with empirical values of eight MIP and six, respectively. It is argued in \cite{Bilki:2014uep} and references therein that these values optimise the selection efficiency in the energy range relevant for the present study.
The event is selected if $5 < i < 15$ to suppress electron contamination and to ensure secondaries that extend over several layers after the interaction.
%\item a cut on minimal number of hits in  two last layers of \ecal\ was applied to reduce number of events with elastic scattering.
\end{itemize}
%This selection scheme originally developed and used in \cite{bib:Naomi}\cite{bib:2012_Doublet}.
%Preselect the events, which have an energy deposition at the last layers of the ECal and interaction in the first half of the Ecal

%---------------------------------------------------------------
%----------------------------TFA--------------------------------
%---------------------------------------------------------------

\section{The \tfa}
\label{sec:track}

%\subsection{Algorithm description}
%The outgoing charged secondary particles in hadron interactions leave tracks in the \ecal. 
%These tracks are reconstructed by a  new simple \tfa\. 
%The resulting observables is used for comparison of different {\sc Geant}4 Monte Carlo simulations with experimental data, taken at FNAL in 2008. 
%Identification of tracks in hadronic showers is useful for precise energy calculation in PFA and for more detailed comparison of experimental data and Monte Carlo simulations. 
%The \tfa\ developed here is oriented on a Monte Carlo comparison study for \ecalp\ using new set of observables, based on properties of secondary particles from hadronic interactions. 

The \tfa\ reconstructs forward-scattered tracks from the interaction between the $\uppi^-$-meson and the absorber material in the absence of a magnetic field.

The algorithm consists of three stages:
\begin{itemize}
\item identification and removal of interaction region;
\item clusterisation of energy deposits;
\item formation of track-like clusters; %and interaction region clusters.
%\item After classification different clusters from a single outgoing secondary particle are merged into one track.
\end{itemize}
The entire algorithm is carried out in the grid units introduced in Sec.~\ref{sec:ecalp}. 
%a unit grid based on the \ecal\ pad identifiers according to 
%pad identifier
%\begin{equation}
%	\vec{x} = (x,y,z)=\left
%\{
%\begin{array}{c}
%x= 0 .. 17 \\ 
%y=0 .. 17 \\
%z = 0 .. 29,
%\end{array}
%\right.
%\end{equation}
%where pad counting starts in the bottom right pad, see Fig.~\ref{fig:ECAL-scheme}. Distances in this grid are measured in {\em grid units}, ${\rm g.u.}$  

%|||||||||||||||||||Interaction zone|||||||||||||||||||||||

\subsection{Identification and removal of the interaction region}
\label{sec:iazone}
A typical inelastic hadronic interaction in the \ecal\ creates a shower with an interaction region and tracks of long-lived particles emerging from it. The interaction region is created by particles such as electrons, photons and low-momentum hadrons that have a short distance of flight in the absorber material of the \ecal. 

In the present analysis the interaction region is defined by all hits that have at least six neighbouring pads with a signal above the noise threshold. 
%These neighbouring pads are also part of the interaction region, which applies in particular at the border of the interaction region. 
For the minimal value of six pads a interaction region is wrongly identified in only 1\% of single muon events. Muon events are used to estimate the fraction of events in which this procedure incorrectly identifies an interaction region.  For the minimal value of six (five) pads, an interaction region is found in 1\% (10\%) of muon events. Increasing the minimal value to seven neighbouring pads with hits further reduces the fraction of events with a fake interaction region but does not alter the results presented below in Sec. 5. 
%The interaction region removal follows a principle similar to that of the isolated hit filter: a hit belongs to the interaction region if it has a number of neighbor hits above a certain threshold. The threshold value of 6 pads is chosen to leave  $\mathrm{\upmu}^-$ events unaffected.
\begin{figure}[H]
\centering
\begin{subfigure}{0.5\textwidth}
\centering
\includegraphics[width=.90\linewidth]{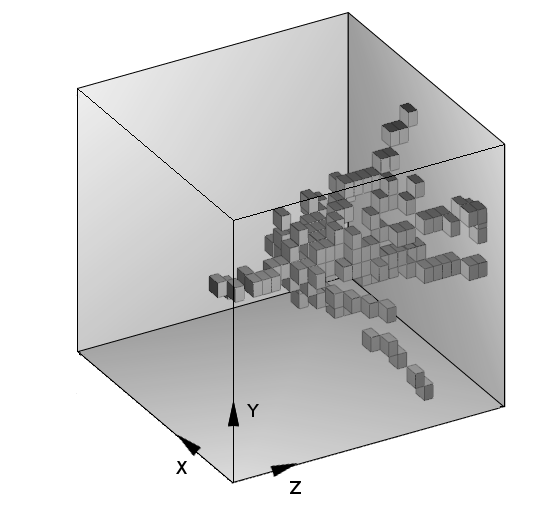}
\caption{\label{fig:before} \sl Before removing the interaction region.}
\end{subfigure}% 
\begin{subfigure}{0.5\textwidth}
\centering
\includegraphics[width=.90\linewidth]{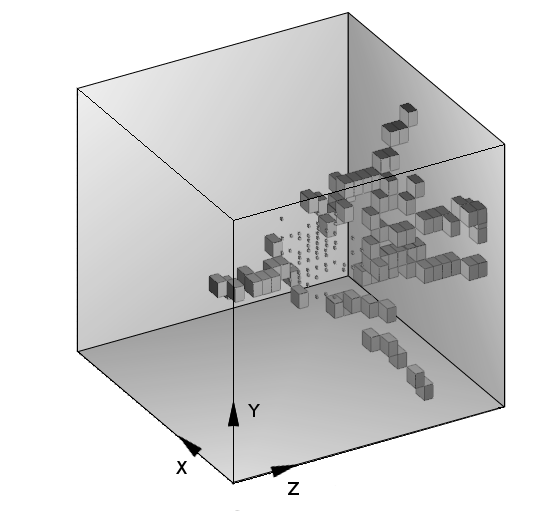}
\caption{\label{fig:after} \sl After removing the interaction region.}
\end{subfigure}
\caption{ \sl Event display of a $\uppi^-$-meson interaction with an energy of 10\,GeV \textit{(a)}  before and \textit{(b)} after removal of the interaction region. Smaller cubes are pads that are part of the interaction region and are not processed by the \tfa . In this event the hits in the first ten layers are classified as hits left by the incoming $\uppi^-$-meson.}

\label{fig:test}
\end{figure}

Figure \ref{fig:before} displays an event after applying noise and isolated hits filters and Fig.\,\ref{fig:after} is the same event after removal of the interaction region, illustrating that the interaction region is the starting point for secondary tracks.

%The energy deposition in the interaction region of the $\mathrm{\uppi}^-$ events in \ecal\ is analyzed in Sec. \ref{sec:results}.

%|||||||||||||||||||||||clusterisation||||||||||||||||||||||||||||

\subsection{Clusterisation of energy deposits}\label{sec:cluster}
During the clusterisation step the energy deposits that are not assigned to the interaction region are grouped into clusters according to topological criteria. %This step is needed to separate secondary tracks into corresponding groups of hits, or clusters, for further classification. 

\begin{figure}[H]
\centering
\includegraphics[width=0.65\textwidth]{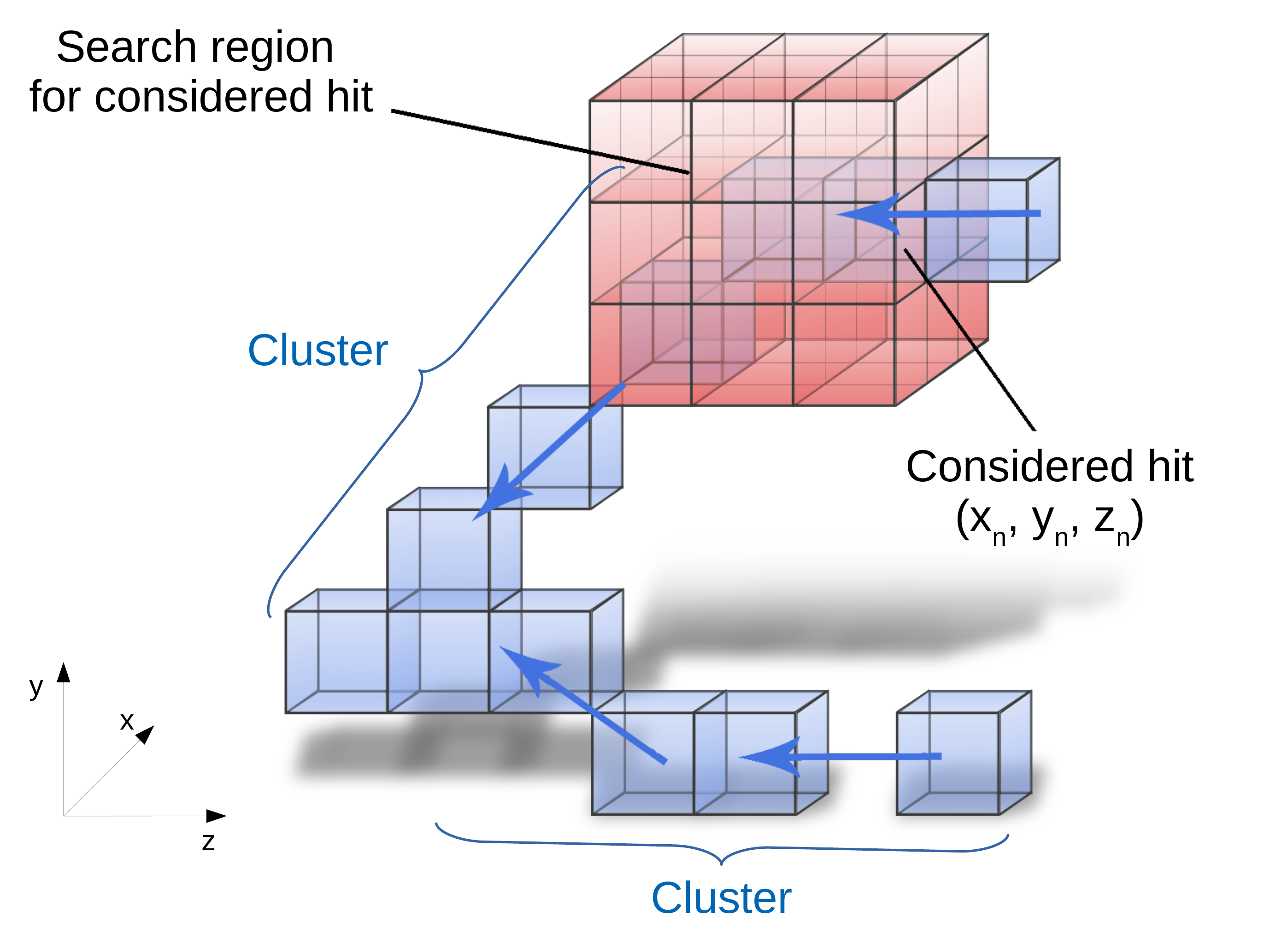}
\caption{\label{fig:democluster} \sl Illustration of the clusterisation step. The \ecal\ hits are represented by blue cubes, and the search region for adjacent hits is indicated by red cubes. The blue arrows point in the direction of the clusterisation flow. }
\end{figure}

%A clusterisation scheme, developed for this study, is called directional recursive clusterisation. 
%The clusterisation algorithm starts recursive assembling of clusters from the last layers of the \ecal, where the secondary tracks are more distinguishable, until the interaction zone or until the first layers in case of MIP-like events. 

The steps of the clusterisation algorithm are described below, with reference to Fig.\,\ref{fig:democluster}.
%his algorithm has the following steps:
\begin{enumerate}
\item The separation of tracks improves with increasing distance from the interaction layer. Therefore, the search for hits to seed a cluster begins in the layers that have largest $z$ and continues in the direction of decreasing $z$. Typically, seeding hits are found in the last layer of the detector;
%For each iteration, the seeding hits for the clusters are taken starting from the last layers of pads to the front layer;
\item A hit can only be attributed to one cluster. This condition excludes double counting of hits. 
A random choice of seeding hits shows that effects arising from ambiguities in the assignment of hits to clusters such as the order in which clusters are created, are negligible;
\item For the clusterisation a nearest-neighbour clustering scheme is applied where for each newly associated hit with coordinates $(x_n,y_n,z_n)$, the algorithm finds nearby hits with the following conditions:
	\begin{itemize}
	\item a neighbour hit should have a $z$ coordinate within $[z_n-2,z_n]$\,g.\,u.\,;
    \item the transverse coordinates of neighbouring hits is searched within ranges $[x_n-1,x_n+1]$ and $[y_n-1,y_n+1]$.
 	\end{itemize}
The search region for nearby hits is visualised  in Fig.~\ref{fig:democluster} as a `red cube' with 3 $\times$ 3 $\times$ 3 pads; 
%with $z$ coordinate equal or less than $z$ coordinate of considered hit
\item  For each newly associated hit the steps 2 and 3 are repeated until the process reaches the first layer of the calorimeter or until no more neighbour hits are found.
\end{enumerate}
The algorithm is motivated by a maximum correspondence between the number of clusters and the number of detected tracks after classification and merging.

%|||||||||||||||||||||||Classification||||||||||||||||||||||||||

\subsection{Formation of tracks}
\label{sec:class}
Secondary long-lived charged particles from hadronic interactions can leave straight, MIP-like tracks in the detector.
The goal of the classification of the clusters obtained in the previous step is thus to select track-like clusters. %and reject clusters left by electromagnetic interaction or detector noise.
%%A track-like cluster is defined as a cluster with hits, spatially arranged in the way, that a trajectory of a single charged particle can go through the most of its pads with energy depositions.

\begin{figure}[H]
\centering
\includegraphics[width=0.5\textwidth]{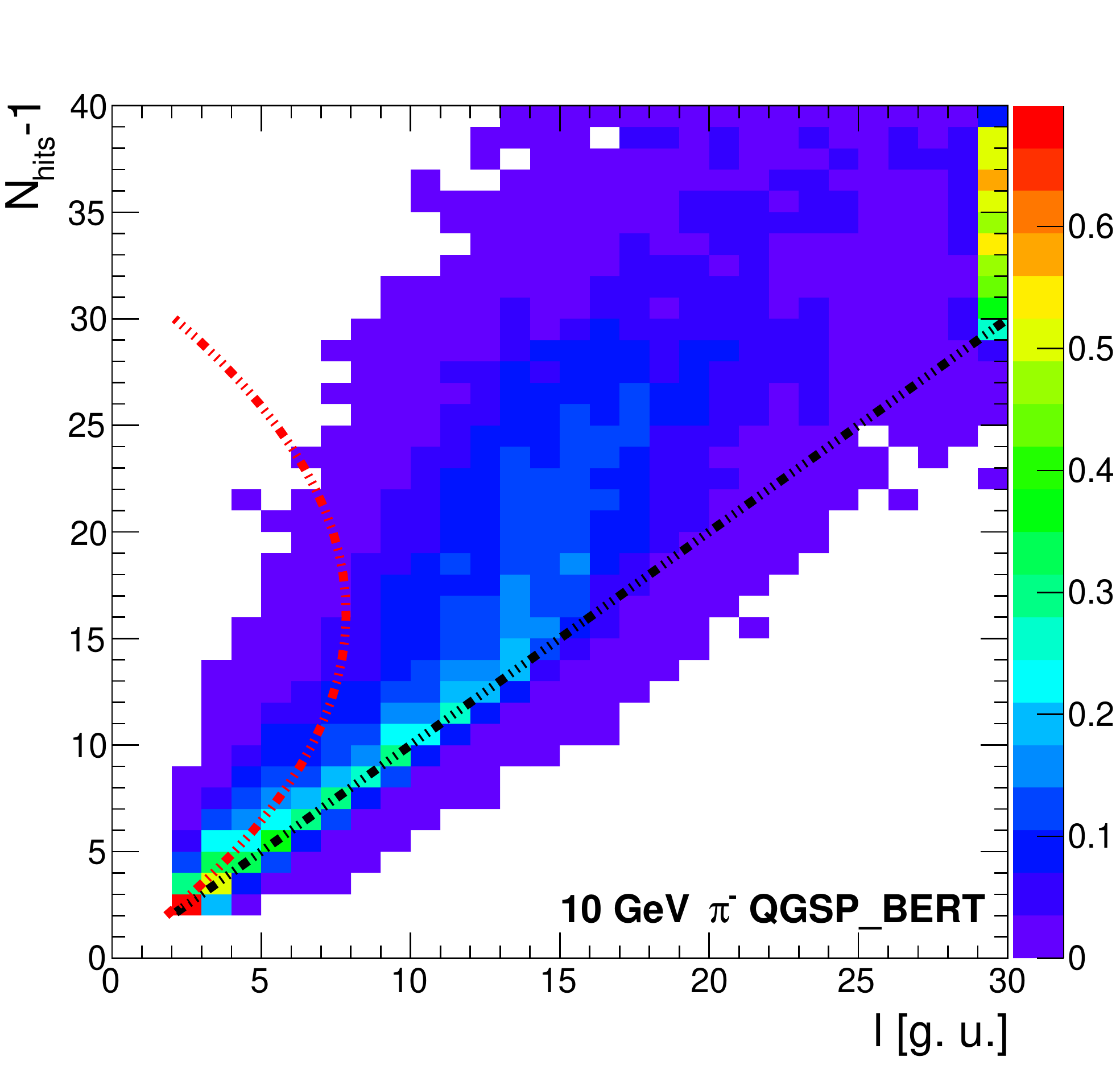}
\caption{\label{fig:lnhits} \sl Correlation between $N_{\mathrm{hits}} - 1$ and cluster length $l$ in ${\sl g.u.}$ for a sample of simulated pion interactions with an energy of 10\,GeV using the \qgsp\ simulation model. Clusters inside the red parabola are rejected by means of Eq.~\ref{eq:observable}. To guide the eye a black line for $N_{\mathrm{hits}} - 1=l$ is also included in the figure.}
\end{figure}

The classification algorithm executes the following steps:
\begin{enumerate}
%\item Reject all clusters with only two hits ($N_{\mathrm{hits}}$) as residual noise clusters;
\item Calculate the number of hits, $N_{\mathrm{hits}}$,  in a cluster and reject all clusters with only two hits as residual noise clusters
\item Calculate the length $l \in \mathbb{R}$ of the considered cluster as the maximal distance between any pair of hits that are in the cluster. For example the lower cluster in Fig.~\ref{fig:democluster} has a length of $\sqrt{\Delta y^2_n +\Delta z^2_n} = \sqrt{25 + 1} \approx 5.1\,{\rm g.u.\,}$;
\item Reject a cluster with a length of less than $l_{\mathrm{cut}} = 2\,{\rm g. u.}$. This corresponds to the minimal length of a track-like cluster with 3 hits; 
\item Compute  the following observable
\begin{equation}
\label{eq:observable}
	\xi = \frac{l}{N_{\mathrm{hits}}-1} + \varepsilon N_{\mathrm{hits}},
\end{equation}
as a measure for the eccentricity of the cluster. 
The first term of Eq.~\ref{eq:observable} is motivated by the linear dependence of $N_{\mathrm{hits}}-1$ on the cluster length $l$, illustrated in Fig.~\ref{fig:lnhits}.
The second term introduces a free parameter $\varepsilon$ as an ad hoc correction to increase the efficiency for selecting clusters that do not match the nominal `pencil-like` topology, as explained below.
The value $\varepsilon=0.03$ was chosen after visual inspection of a few tens of events for pion energies of 10\,GeV in an event display. 
The chosen value is a compromise between a too small value at which also muon tracks would get assigned more than one track and too large values at which even for electrons the algorithm would result in one single track.
The choice made for 10\,GeV is also adequate for the other energies relevant for this analysis. For a detailed discussion see Refs.~\cite{Poschl:2016ehb} and~\cite{bilokin:tel-01826535} ;
%%In Eq. \ref{eq:observable} $\varepsilon$ is a free parameter to correct for non-ideal pencil like clusters. 
%%The parameter $\varepsilon$ takes values much smaller than 1. 
\item If $\xi \geq 1$, a cluster is considered as track-like. Otherwise, the cluster is classified as two inseparable tracks.
\end{enumerate}
Due to effects such as multiple scattering, residual detector-noise, $\delta$-rays or the residual arbitrariness in the assignment of hits to clusters, the reconstructed tracks are in general not exactly pencil-like.  
The correction term $\varepsilon N_{\mathrm{hits}}$ in the definition of $\xi$ serves to keep a cluster as track-like even if it has large $N_{\mathrm{hits}}$ and its form is not strictly pencil-shaped, i.e. $l / (N_{\mathrm{hits}}-1) < 1$. 

%\begin{figure}[H]
%\centering
%\includegraphics[width=0.5\textwidth]{stdselection/muon.png}
%\caption{\label{fig:demomerge} \sl An example of a segmented track in the \ecal\ from a single $\upmu^-$.}
%\end{figure}

%The next step of the classification is to detect a cluster from the primary particle. 
%If this cluster exists and meets the conditions for a track-like cluster, it affects the track counting and merging algorithm.
A cluster is classified as being produced by the incoming $\uppi^-$-meson if it starts in the first module of the \ecal\ and if it has a small polar angle with respect to the $z$-axis. 
An example of a cluster produced by an incoming $\uppi^-$-meson is visible in Fig. \ref{fig:after}. Clusters assigned to the incoming $\uppi^-$-meson are discarded in the following analysis. The remaining track-like clusters are merged into tracks if the relative angle $\theta_\mathrm{c}$ between these clusters fulfils the condition $\sin \theta_{\mathrm c} < 0.15$, optimised using the polar angle distribution of secondary particles in simulated $\uppi^-$-meson interactions.
%by the comparison of outgoing particles and outgoing tracks after event generation and simulation, respectively.

%Different track-like clusters that correspond to track segments from a single particle, see Fig. \ref{fig:demomerge}, have to be merged into a single track. 
%The merging procedure combines track-like clusters with any type of clusters using a simple cone algorithm {\bf cone size?}.
%Because of limitations of clusterisation method, 
%A segmented track from a secondary particle can be divided into two different clusters as it is shown in Fig. \ref{fig:demomerge}.
%It is thus necessary to merge different track segments from a single particle into one track-like cluster. 
%The merging procedure combines track-like clusters with any type of clusters using a simple cone algorithm. 
%TODO: extend merging?
For a sample of single, isolated 6 GeV muons, the track-finding algorithm finds a single track with a 99.7\% efficiency. 
%Tested on a sample of single, isolated muons with an energy of 6\,GeV simulated with the \ftfp\ physics list, the \tfa\ finds correctly only one track with 99.7\% efficiency. The sample contains about 3\% events with segmented tracks. 

%------------------------------------------------------------
%----------------------SYSTEMATICS---------------------------
%------------------------------------------------------------

%------------------------------------------------------------------
%---------------------------RESULTS--------------------------------
%------------------------------------------------------------------

\section{Results}
%Systematic uncertainties related to the calorimeter calibration, event selection and preprocessing are discussed in Ref. \cite{bib:Naomi}.
\label{sec:results}
%\subsection{Comparing Monte Carlo models with data}
%Various Monte Carlo models are compared with the test beam data in terms of the interaction region parameters, and secondary tracks observables. The following figures show these comparisons for simulations based on the two studied Monte Carlo physics lists.
Observables characterising the interaction region and secondary particles are measured in data and are compared with predictions of the three \geantfour\ simulation models introduced in Sec.\,\ref{mc-models}. The average values of observables are also used to make quantitative comparisons.
After pre-selection the data have a residual contamination of 8.8\% (1.5\%) double $\uppi^-$-meson events at 2\,GeV (10\,GeV) beam energy~\cite{Bilki:2014uep}.  Therefore, for comparison with data all simulation samples were produced with an admixture of double $\uppi^-$ events. For average values of observables, correction factors are extracted for each Monte Carlo sample by comparing the results of contaminated samples with those from pure samples. To account for residual contamination, the averages of the data are multiplied by {\em final} correction factors. This final correction factors are given by the arithmetic mean of correction factors of the three considered simulation models. The final correction factors are between 0.99 and 1.01 and their uncertainties are much less than one percent. 
%For each data point and each physics list correction factors are calculated and the averages of these correction factors are the final correction factors. 
%The systematic uncertainties of the final correction factors are determined from the difference between these correction factors and the individual correction factors. 
The total systematic error never exceeds two percent and is dominated by two other sources of systematic uncertainties that have been studied.
These are the lowering of the MIP threshold from the nominal 0.6 to 0.4 and the uncertainty on the absolute MIP energy scale~\cite{Adloff:2013vra} that has been varied by $\pm2\%$. 
%The resulting total systematic uncertainties from these sources vary mainly between 0.5\% and 1.5\%.
%These variations lead to minor changes on the results and are neglected in the following. 
%Another source of systematic uncertainty suggested in~\cite{Adloff:2013vra} that may be caused by  has been studied and was found to be negligible for the results presented in the following.
%Systematic uncertainties related to the calibration, event selection and preprocessing are discussed in Ref. \cite{bib:Naomi}. 

\subsection{Energy fraction of the interaction region}
%---------------interaction zone--------------------
The first estimator to characterise the interaction of $\uppi^-$-mesons with the absorber material is the fraction 
\begin{equation}
f_{\mathrm{IR}} = \frac{E_{\mathrm{IR}}}{E_{\mathrm{tot}}}
\end{equation}
where $E_{\mathrm{IR}}$ is the energy deposited in the interaction region and $E_{\mathrm{tot}}$ is the total energy deposited in the \ecal\ .

%In this study the interaction region created by $\uppi^-$ interacting with the absorber material is characterised by the fraction $f_{IR}$ of total energy deposited in the calorimeter.% and its lateral radius $r_{IR}$ averaged over hits.

\begin{figure}[H]
\centering
\begin{subfigure}{0.5\textwidth}
\centering
\includegraphics[width=.90\linewidth]{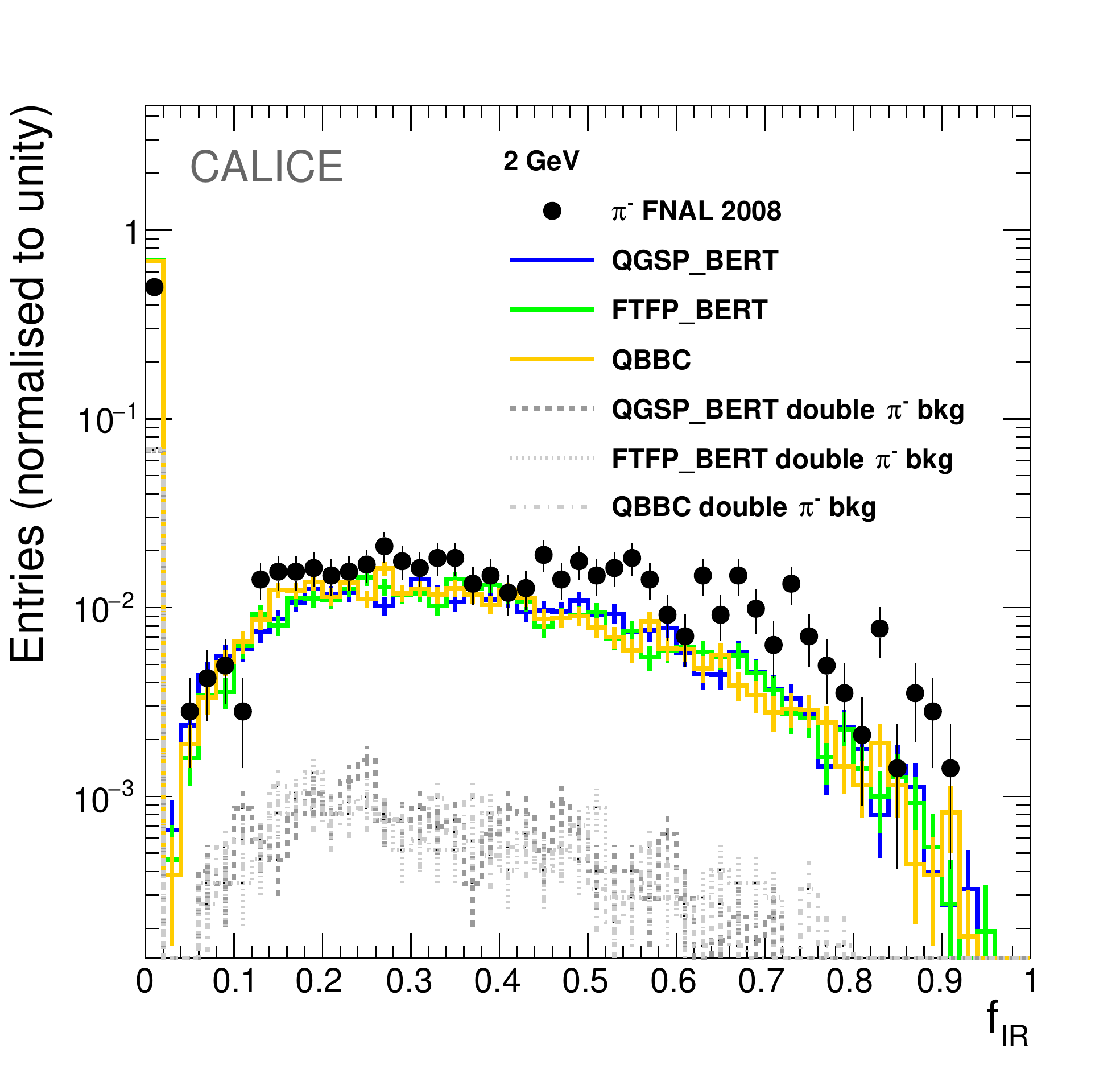}
\caption{\label{fig:efr2} }
\end{subfigure}% 
\begin{subfigure}{0.5\textwidth}
\centering
\includegraphics[width=.90\linewidth]{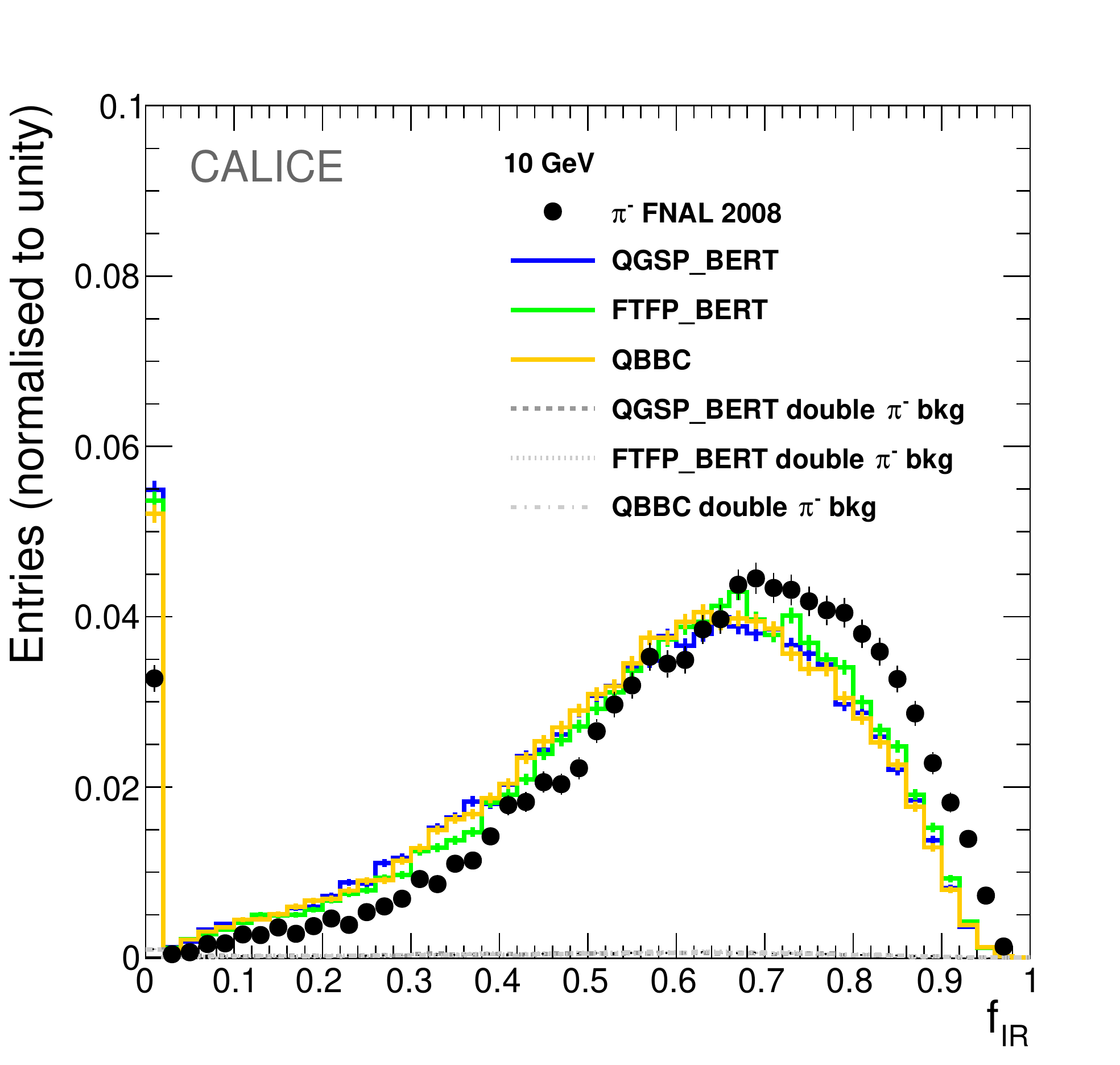}
\caption{\label{fig:efr10} }
\end{subfigure}
\caption{\label{fig:irexample} \sl The $f_{\mathrm{IR}}$ distribution for energies of 2 GeV (a) and 10 GeV (b) of the beam energy as observed in data (points with error bars) and for the three simulation models, \qgsp , \ftfp\ and \qbbc . The double $\uppi^-$-meson background for each of the three models is also shown. The first bin contains events without a detected interaction region. All histograms are normalised to unit area. Error bars represent statistical uncertainties only.
}
\end{figure}

Figure~\ref{fig:irexample} compares the distribution of $f_{\mathrm{IR}}$ in data with the three simulation models. The lowest bin of these histograms corresponds to the fraction of events for which no interaction region is found by the algorithm.
%%As can be seen from Fig. \ref{fig:irexample} the fraction of events in this bin for the chosen physics list is approximately(?) compatible with data. 
The rest of the distribution can be approximately described by a skewed normal distribution. The mean value of $f_{\mathrm{IR}}$ is shifted towards larger values in data with respect to simulation. Qualitatively, this observation suggests a different repartition of the deposited energy between the dense interaction region and the sparser parts of the shower in data and the simulation models.

\begin{figure}[H]
\centering
\includegraphics[width=0.5\textwidth]{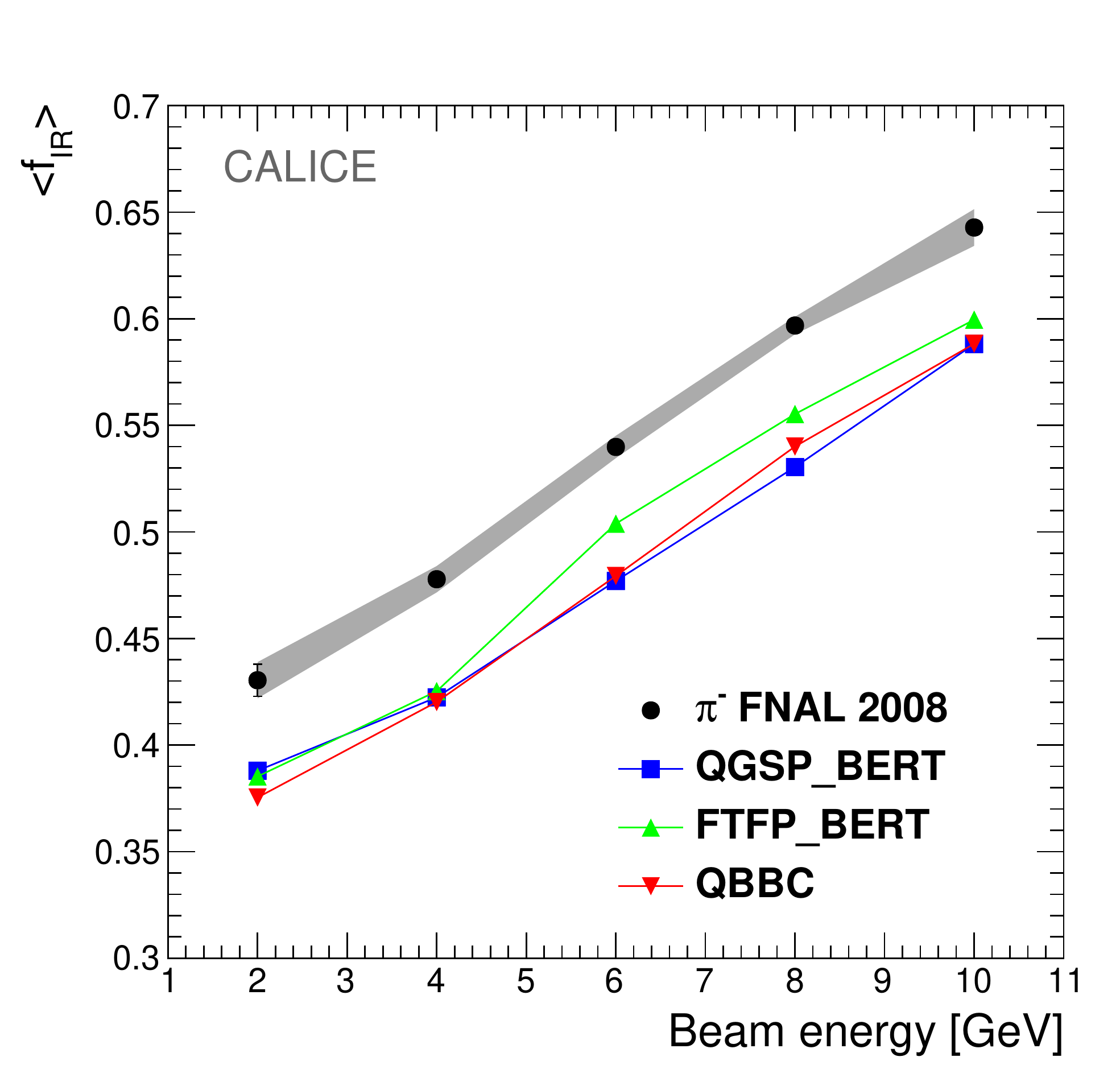}
\caption{\label{fig:irgraph} \sl The average fraction $\left< f_{\mathrm{IR}}\right>$ as a function of the beam energy for data (black points with grey shaded error band) and the three simulation models. 
%\qgsp\ (blue squares), \ftfp\ (green upward-pointing triangles) and \qbbc\ (red downward-pointing triangles). 
Error bars represent statistical errors. The error band is the sum in quadrature of the systematic error and the statistical error.  
Only events for which an interaction region has been detected are included in $\left< f_{\mathrm{IR}}\right>$.}
%Events without a detected interaction region according to Sec.~\ref{sec:iazone} are discarded.}
\end{figure}

Figure~\ref{fig:irgraph} shows the average value of $f_{\mathrm{IR}}$,  $\left< f_{\mathrm{IR}}\right>$, as a function of beam energy  for beam energies of 2, 4, 6, 8 and 10\,GeV. 
Only events in which an interaction region has been detected are included in $\left< f_{\mathrm{IR}}\right>$. 
%Events without a detected interaction region according to Sec.~\ref{sec:iazone} are discarded. 
An increase of $\left< f_{\mathrm{IR}}\right>$  with increasing beam energy from 43\% to around 64\% is observed. Qualitatively, this is expected as number of particles increases with increasing energy but also the electromagnetic component of the hadronic shower becomes increasingly important for higher energies of the interacting $\uppi^-$-meson. All three simulation models underestimate the energy fraction by about 10--15\% while the slope is reproduced to a much better level. 
%In Fig. \ref{fig:fulliregraph} the mean value of $f_{IR}$ is shown as a function of the beam energy  for beam energies of 2, 4, 6, 8 and 10\,GeV. Events without a detected interaction region according to Sec.~\ref{sec:iazone} are discarded. An increase of $f_{IR}$ with increasing beam energy from 43\% at 2\,GeV to around 64\% around is observed. Qualitatively this is expected as the electromagnetic component of the hadronic shower becomes increasingly dominant with increasing energy of the primary particle.
%In case of the \geantfour\ Version 10.1, see Fig~\ref{fig:iregraph10}, the mean value is consistently about 10-15\% smaller than observed in the data for all physics lists.  The change of slope of the \ftfp\ physics above 4\,GeV, i.e. at the sharp transition between the Bertini cascade and the Fritiof model, is less prominent in Version 10.1 than in Version 9.6 as demonstrated by Fig.~\ref{fig:iregraph96}, see also Ref.~\cite{bib:can-055}.  A change in shape is also visible for the physics list \qgsp\ modifying the, loosely speaking, parabola shape closer to a straight line, which is at least true above 4\,GeV.
%consistent with the underestimation of the total energy deposition by the models reported in~\cite{Bilki:2014uep}.

%\begin{figure}[H]
%\centering
%\begin{subfigure}{0.5\textwidth}
%\centering
%\includegraphics[width=.90\linewidth]{stdselection/e-ir-graph.pdf}
%\caption{\label{fig:iregraph10} }
%\end{subfigure}% 
%\begin{subfigure}{0.5\textwidth}
%\centering
%\includegraphics[width=.90\linewidth]{stdselection/e-ir-graph-v96.pdf}
%\caption{\label{fig:iregraph96} }
%\end{subfigure}
%\caption{\label{fig:fulliregraph} \sl  
%The average fraction $\left< f_{IR}\right>$ as as function of the beam energy for data (black points with error shaded band) in comparison to the three {\sc Geant4} physics lists, \qgsp\ ( blue squares), \ftfp\ (green upward-pointing triangles) and \qbbc\ (red downward-pointing triangles) generated with version 10.1 (a) and in comparison to the \qgsp\ and \ftfp\ models generated with version 9.6 (b).  Error bars represent statistical errors and the error band the systematic error from the correction for double $\uppi$ events. 
%{\bf think about removing the following} Events without a detected interaction region according to Sec.~\ref{sec:iazone} are discarded.
%}
%\end{figure}

\subsection{Lateral radius of interaction region}\label{sec:latrad}
\begin{figure}[H]
\centering
\begin{subfigure}{0.5\textwidth}
\centering
\includegraphics[width=.90\linewidth]{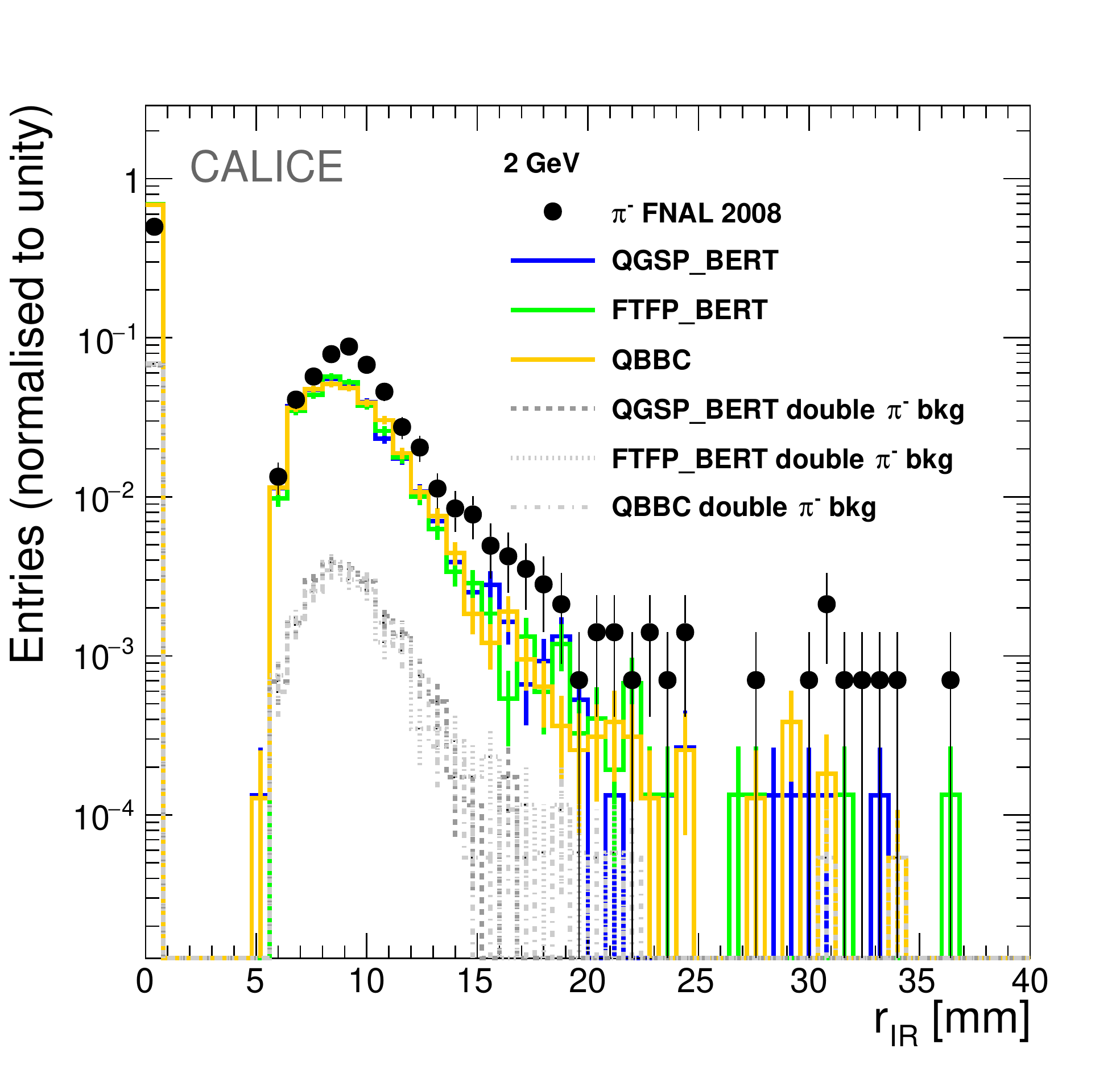}
\caption{\label{fig:rir2} }
\end{subfigure}% 
\begin{subfigure}{0.5\textwidth}
\centering
\includegraphics[width=.90\linewidth]{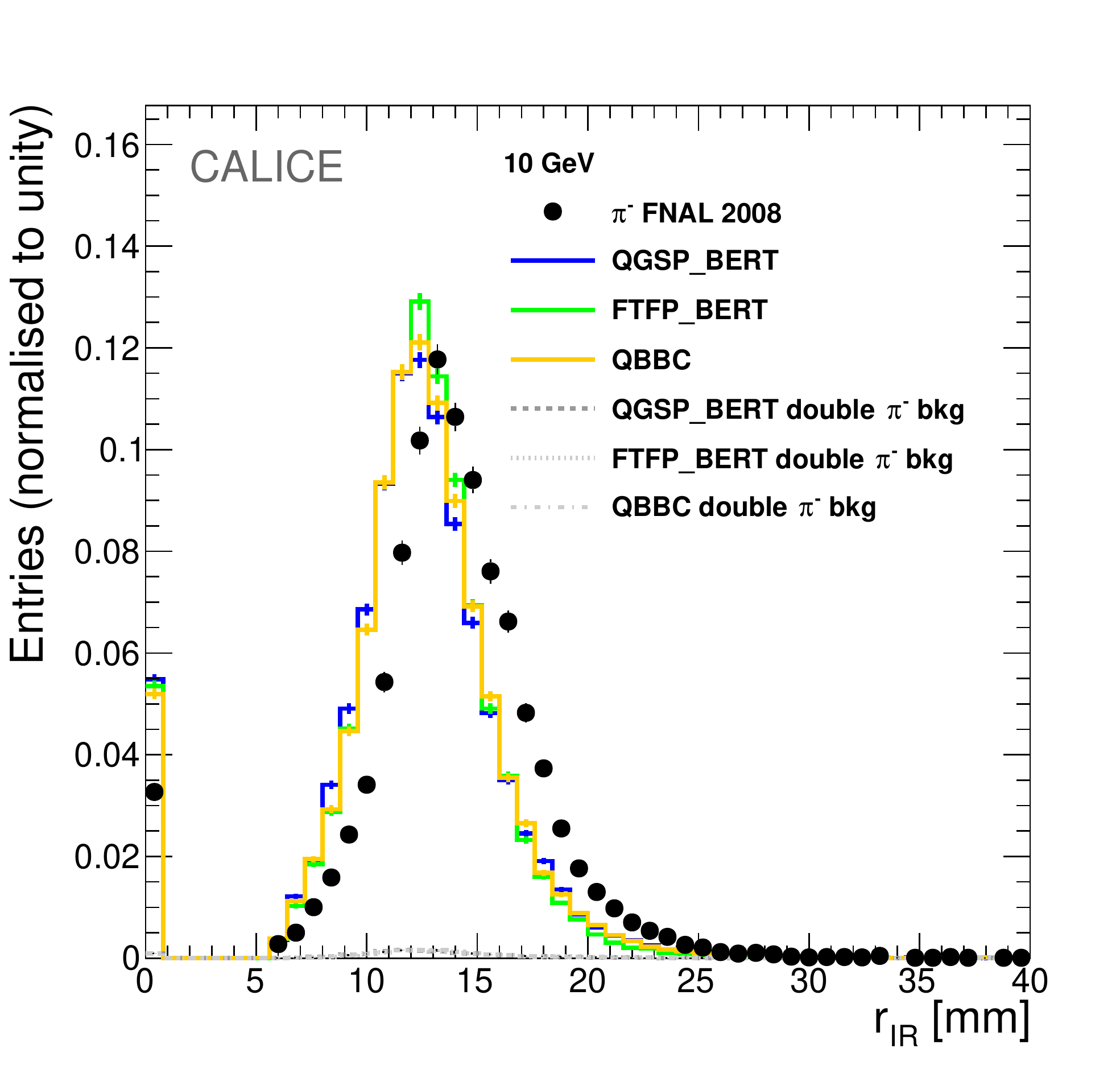}
\caption{\label{fig:rir10} }
\end{subfigure}
\caption{\label{fig:rirexample} \sl The $r_{\mathrm{IR}}$ distribution for energies of 2 GeV (a) and 10 GeV (b) of the beam energy. Other details follow those of Fig.~\ref{fig:irexample}. 
%Furthermore, the same description as for Fig.~\ref{fig:irexample} applies.  
%as observed in data (points with error bars) and for the three simulation models, \qgsp\ (blue histogram), \ftfp\ (green histogram) and \qbbc\ (yellow histogram).  The double $\uppi^-$ background for the three models is shown by the grey dashed, dotted and dash-dotted histograms. The first bin contains events without a detected interaction region. All histograms are normalised to unity. Error bars represent statistical uncertainties only.
}
\end{figure}

\begin{figure}[H]
\centering
\includegraphics[width=0.5\textwidth]{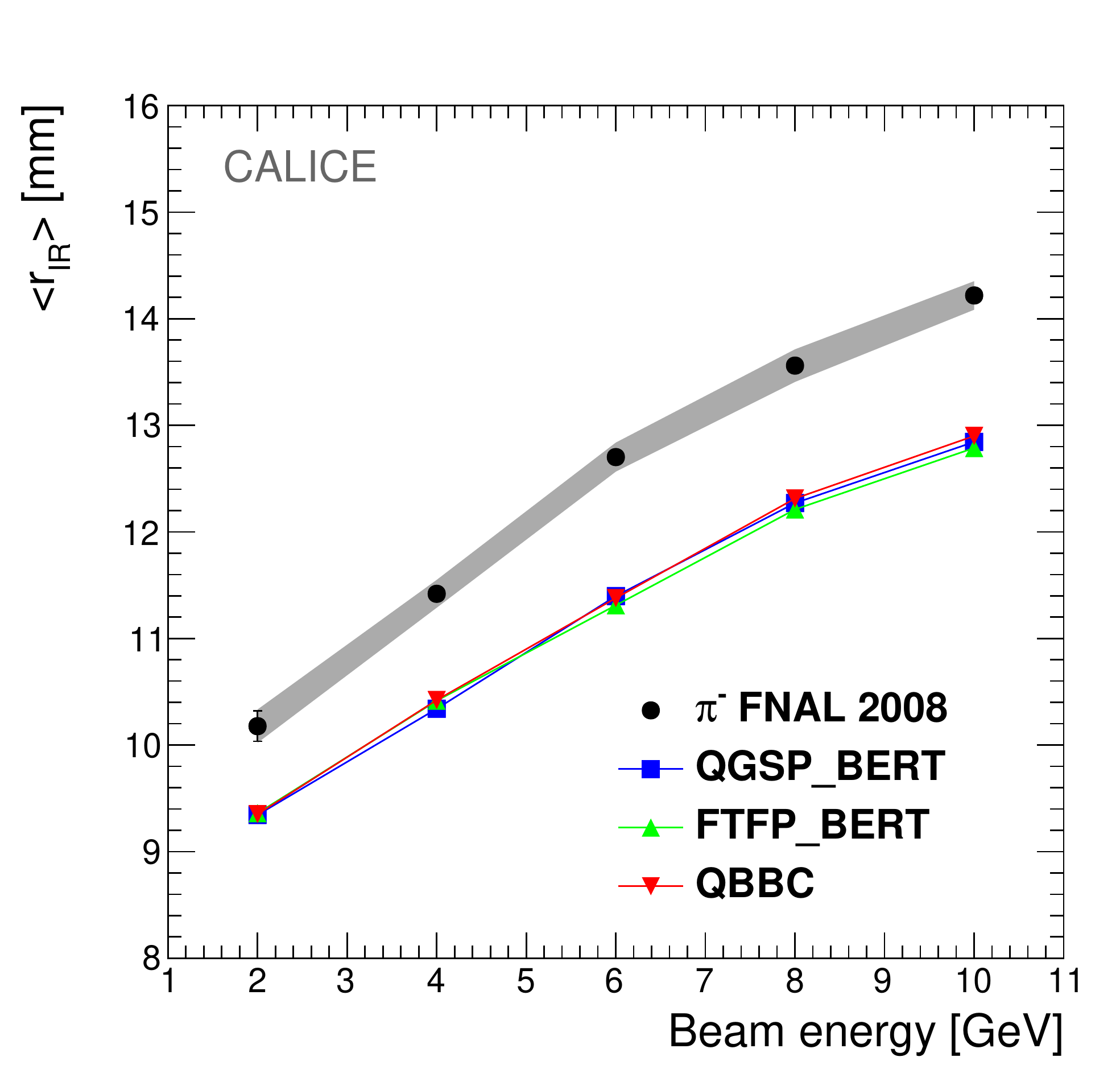}
\caption{\label{fig:irrgraph} \sl The average radius $\left< r_{\mathrm{IR}} \right>$ as a function of the beam energy. Other details follow those of Fig.~\ref{fig:irgraph}. 
%Furthermore, the same description as for Fig.~\ref{fig:irgraph} applies. 
%for data (black points with error shaded band) in comparison to the three simulation models, \qgsp\ (blue squares), \ftfp\ (green upward-pointing triangles) and \qbbc\ (red downward-pointing triangles). Error bars represent statistical errors and the error band the systematic error from the correction for double $\uppi^-$ events. Events without a detected interaction region according to Sec.~\ref{sec:iazone} are discarded.
}
\end{figure}

%|||||||||||||||Radius of interaction zone|||||||||||||||||||
%Along with $f_{IR}$, 
The lateral radius $r_{\mathrm{IR}}$  of the detected interaction region is a measure of the spatial extension of the interaction region. It is defined as:
\begin{equation}
%r_{IR} = \frac{\displaystyle \sum_{hit \in IR} \sqrt{(\bar{x}_{IR} - x_{hit})^2 + (\bar{y}_{IR} - y_{hit})^2}} {\displaystyle N_{hits}^{IR}},
r_{\mathrm{IR}} =  \displaystyle \frac{1}{N_{\mathrm{hits}}^{\mathrm{IR}}}  \displaystyle \sum_{\mathrm{hit} \in \mathrm{IR}} \sqrt{(\bar{x}_{\mathrm{IR}} - x_{\mathrm{hit}})^2 + (\bar{y}_{\mathrm{IR}} - y_{\mathrm{hit}})^2}\,\,\,,
\label{eq:rir}
\end{equation}
where the sum runs over the hits in the interaction region, here labelled by $\mathrm{IR}$, and $N_{\mathrm{hits}}^{IR}$ is the number of hits in the interaction region. In Eq.~\ref{eq:rir} $\bar{x}_{\mathrm{IR}}$ and $\bar{y}_{\mathrm{IR}}$ are the transverse coordinates of the barycentre of the interaction region, which in analogy with Eq.~\ref{eq:barycentre}, are defined as: 
\begin{eqnarray}
\label{eq:barycentre2}
	\bar{x}_{\mathrm{IR}} = \frac{\displaystyle \sum_{\mathrm{hit} \in \mathrm{IR}} x_{\mathrm{hit}}\,E_{\mathrm{hit}}}{\displaystyle \sum_{\mathrm{hit} \in \mathrm{IR}} E_{\mathrm{hit}}} 
    \text{\,\,\,\,\,\,and\,\,\,\,\,\,}
   	\bar{y}_{\mathrm{IR}} = \frac{\displaystyle \sum_{\mathrm{hit} \in \mathrm{IR}} y_{\mathrm{hit}}\,E_{\mathrm{hit}}}{\displaystyle \sum_{\mathrm{hit} \in \mathrm{IR}} E_{\mathrm{hit}}}.
\end{eqnarray}
%15 to suppress electron contamination and to ensure ‘long‘ secondary tracks
%127
%after the interaction.
%128
%4    The track-finding algorithm
%129
%The track-finding algorithm reconstructs forward-scattered tracks from the interaction between the primary
%1
Distributions of $r_{\mathrm{IR}}$ for data and the predictions of the three simulation models are displayed in Fig.~\ref{fig:rirexample}
for $\uppi^-$-meson energies of 2 and 10\,GeV. In both cases, the measured interaction region is wider than the predictions by the simulation models.  
Figure \ref{fig:irrgraph} displays the dependence of the average $r_{\mathrm{IR}}$, $\left<r_{\mathrm{IR}}\right>$, on the beam energy for the data and the three simulation models. Again, only events in which an interaction region has been detected are included in $\left< r_{\mathrm{IR}} \right>$. The lateral size of the interaction region increases with increasing beam energy. This trend is the same for data and for the simulation models. The interaction region measured in data is about 10\% wider than those predicted by the three simulation models, all of which yield similar distributions. 
%For all tested energies the interaction region measured in data is constantly around 10\% wider than is the case of the \geantfour\ physics lists that lead to identical results. 
%An interpretation of this observation may be that the simulation is lacking energy depositions by secondaries with a comparatively long mean free path length. 

%|||||||||||||||||||Number of clusters||||||||||||||||||||||
\subsection{Number of clusters}
\begin{figure}[H]
\centering
\begin{subfigure}{0.5\textwidth}
\centering
\includegraphics[width=.90\linewidth]{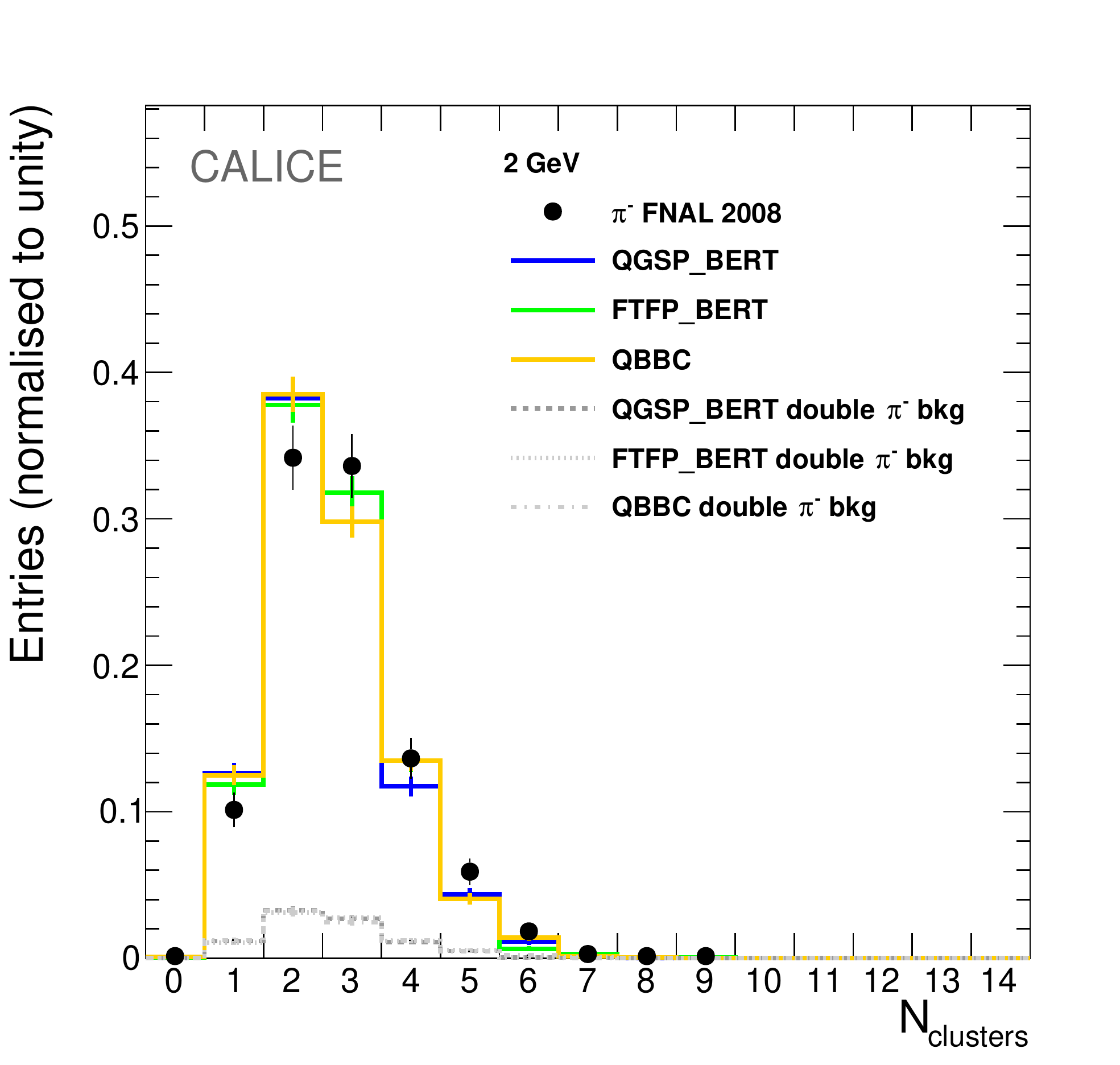}
\caption{\label{fig:cl2} }
\end{subfigure}% 
\begin{subfigure}{0.5\textwidth}
\centering
\includegraphics[width=.90\linewidth]{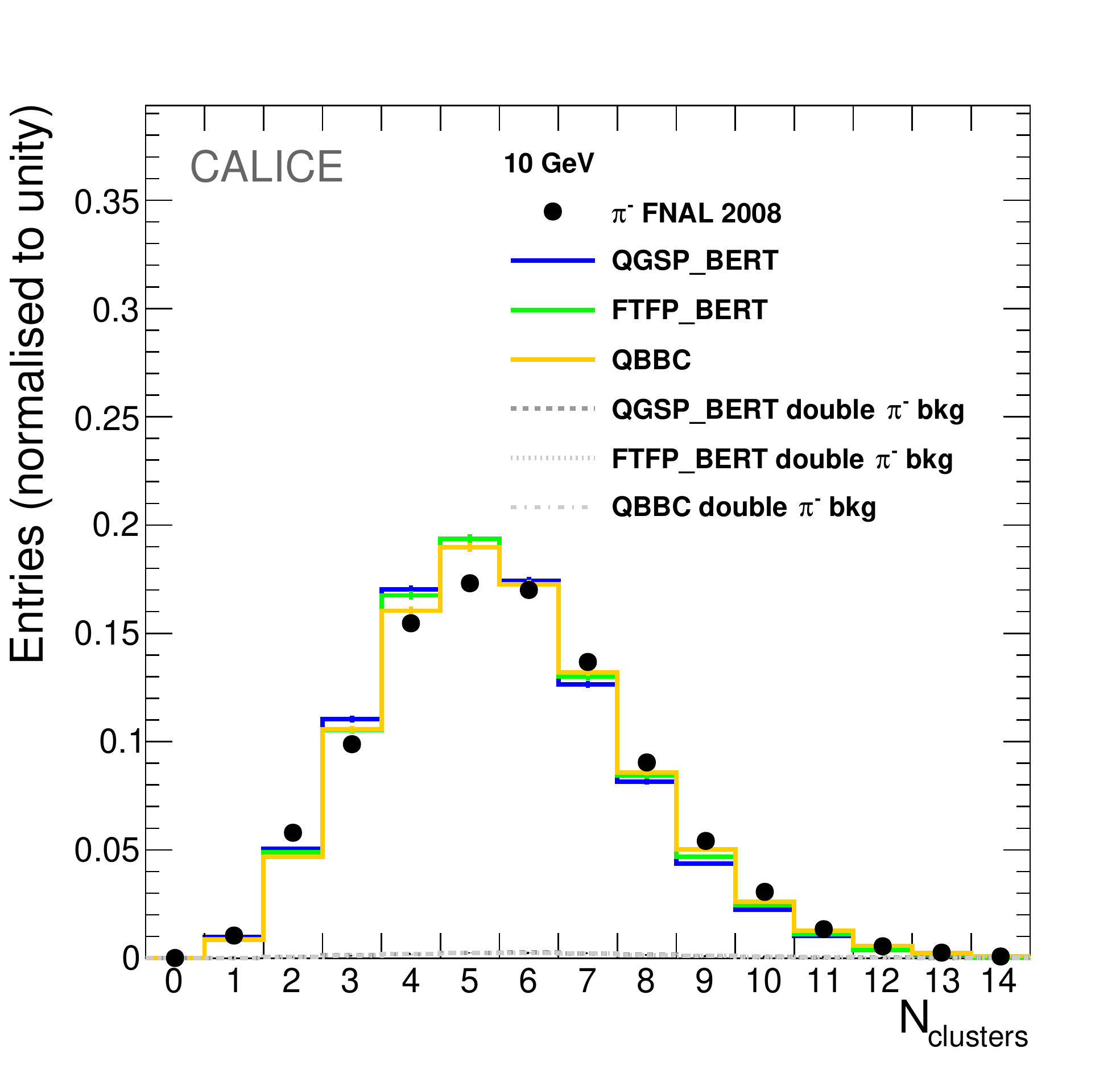}
\caption{\label{fig:cl10} }
\end{subfigure}
\caption{\label{fig:clusterexample} \sl Number of clusters for energies of 2 GeV (a) and 10 GeV (b) of the beam energy. Only events for which an interaction region has been detected have been included.  Other details follow those of Fig.~\ref{fig:irexample}.
%Furthermore, the same description as for Fig.~\ref{fig:irexample} applies but events without a detected interaction region according to Sec.~\ref{sec:iazone} are discarded.
%observed in data (points with error bars) and for the three simulation models, \qgsp\ (blue histogram), \ftfp\ (green histogram) and \qbbc\ (yellow histogram)The double $\uppi^-$ background for the three models is shown by the grey dashed, dotted and dash-dotted histograms. All histograms are normalised to unity. Error bars represent statistical uncertainties only.
}
\end{figure}

The final tracks are composed from segments that are given by clusters, as described in Sec.\,\ref{sec:cluster}. This motivates studying the total number of clusters, $N_{\mathrm{clusters}}$, reconstructed in each event by the \tfa. This observable is independent of details of the \tfa\, since it depends neither on the \ep\ value nor on other free parameters of the classification algorithm. Note that in all following discussion, only events in which an interaction region has been detected are considered.
Figure~\ref{fig:clusterexample} compares the distribution of $N_{\mathrm{clusters}}$ in data with the predictions of the three \geantfour\ simulation models for incoming $\uppi^-$-mesons with energies of 2 and 10\,GeV. The data are described well by the simulation albeit being slightly shifted towards higher values.

%The $N_{clusters}$ distribution is given in Fig.~\ref{fig:clusterexample} for data and the three \geantfour\ simulation models for energies of 2 and 10\,GeV of the incoming $\uppi^-$-meson, respectively. %While the data and MC agree in many bins within errors, the distributions are nevertheless slightly shifted w.r.t. each other.  

\begin{figure}[H]
\centering
\includegraphics[width=0.5\textwidth]{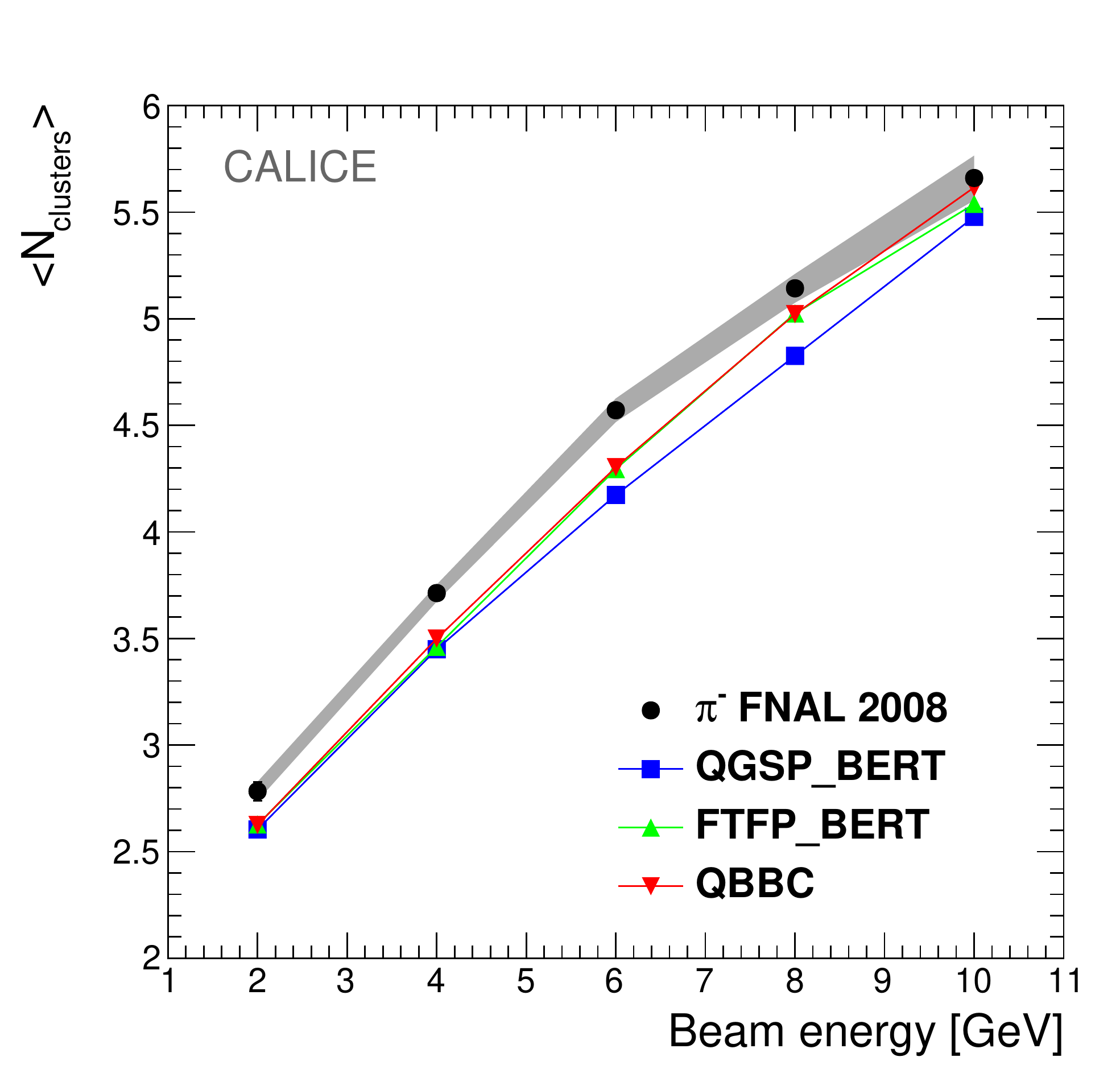}
\caption{\label{fig:clustergraph} \sl The average number of clusters $\left< N_{\mathrm{clusters}} \right>$ as a function of the beam energy. Other details follow those of Fig.~\ref{fig:irgraph}.  
%for data (black points with error shaded band) in comparison to the three simulation models, \qgsp\ (blue squares), \ftfp\ (green upward-pointing triangles) and \qbbc\ (red downward-pointing triangles). Error bars represent statistical errors and the error band the systematic error from the correction for double $\uppi$ events.
}
\end{figure}

Figure \ref{fig:clustergraph} shows the dependence of the average number of clusters, $\left<N_{\mathrm{clusters}}\right>$, on the beam energy for data and the simulation models. The predictions of the models are systematically below data at all energies. The largest deviation is about 7\%. The agreement tends to improve with increasing beam energy and is best at 10\,GeV. 
%At 2 and 10\,GeV there is a good agreement between data and both simulation samples, but for intermediate beam energies the \tfa\ tends to find more clusters in data than in Monte Carlo.

%||||||||||||||||||||Number of tracks||||||||||||||||||||||||
\subsection{Number of tracks} \label{sec:ntracks}
\begin{figure}[H]
\centering
\begin{subfigure}{0.5\textwidth}
\centering
\includegraphics[width=.90\linewidth]{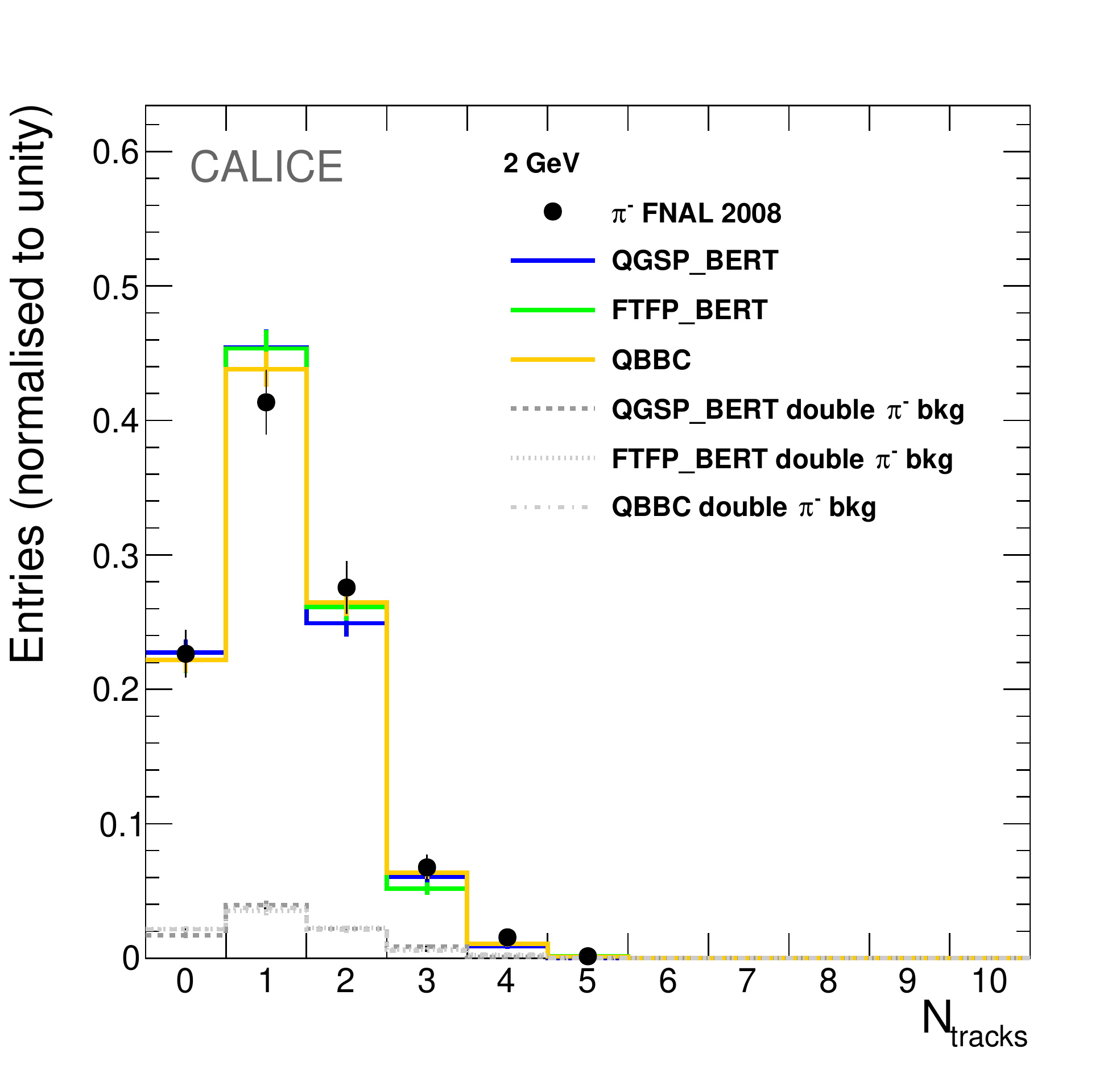}
\caption{\label{fig:tr2} }
\end{subfigure}% 
\begin{subfigure}{0.5\textwidth}
\centering
\includegraphics[width=.90\linewidth]{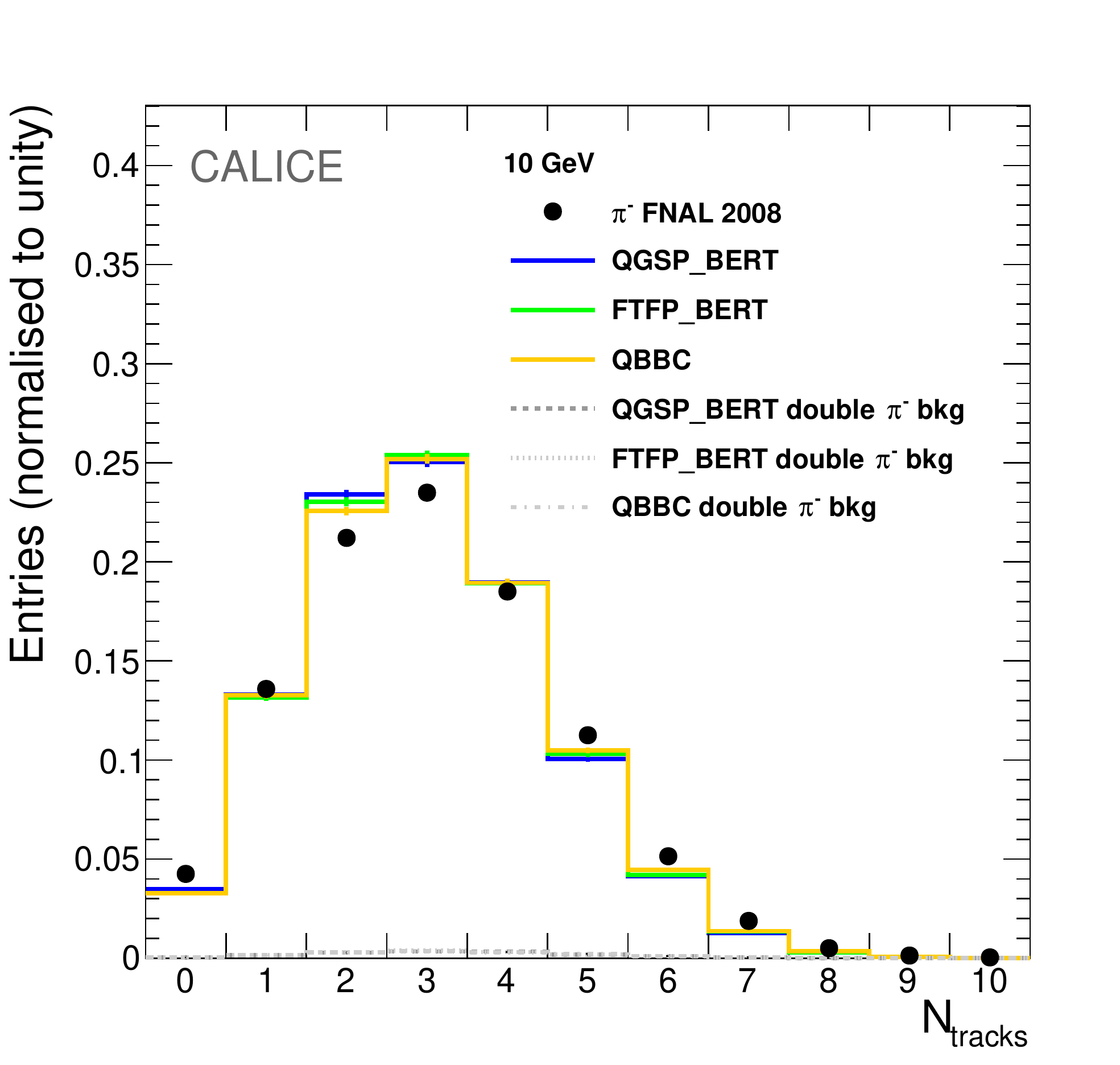}
\caption{\label{fig:tr10} }
\end{subfigure}
\caption{\label{fig:trackexample} \sl Number of secondary tracks for energies of 2 GeV (a) and 10 GeV (b) of the beam energy. Other details follow those of Fig.~\ref{fig:clusterexample}.  
%as observed in data (points with error bars) and for the three models, \qgsp\ (blue histogram), \ftfp\ (green histogram) and \qbbc\ (yellow histogram) The double $\uppi^-$ background for the three models is shown by the grey dashed, dotted and dash-dotted histograms. All histograms are normalised to unity. Error bars represent statistical uncertainties only.
}
\end{figure}

%A central result of the \tfa\ is the number of secondary tracks ($N_{\mathrm{tracks}}$) and observables based on their properties.
The $N_{\mathrm{tracks}}$ distributions are given in Fig.~\ref{fig:trackexample} for data and the three simulation models for energies of 2 and 10\,GeV of the incoming $\uppi^-$-mesons, respectively. 
Data and simulation are in good agreement,  although at 10\,GeV the simulation predicts a narrower spread in $N_{\mathrm{tracks}}$ than data. 

\begin{figure}[H]
\centering
\includegraphics[width=.50\linewidth]{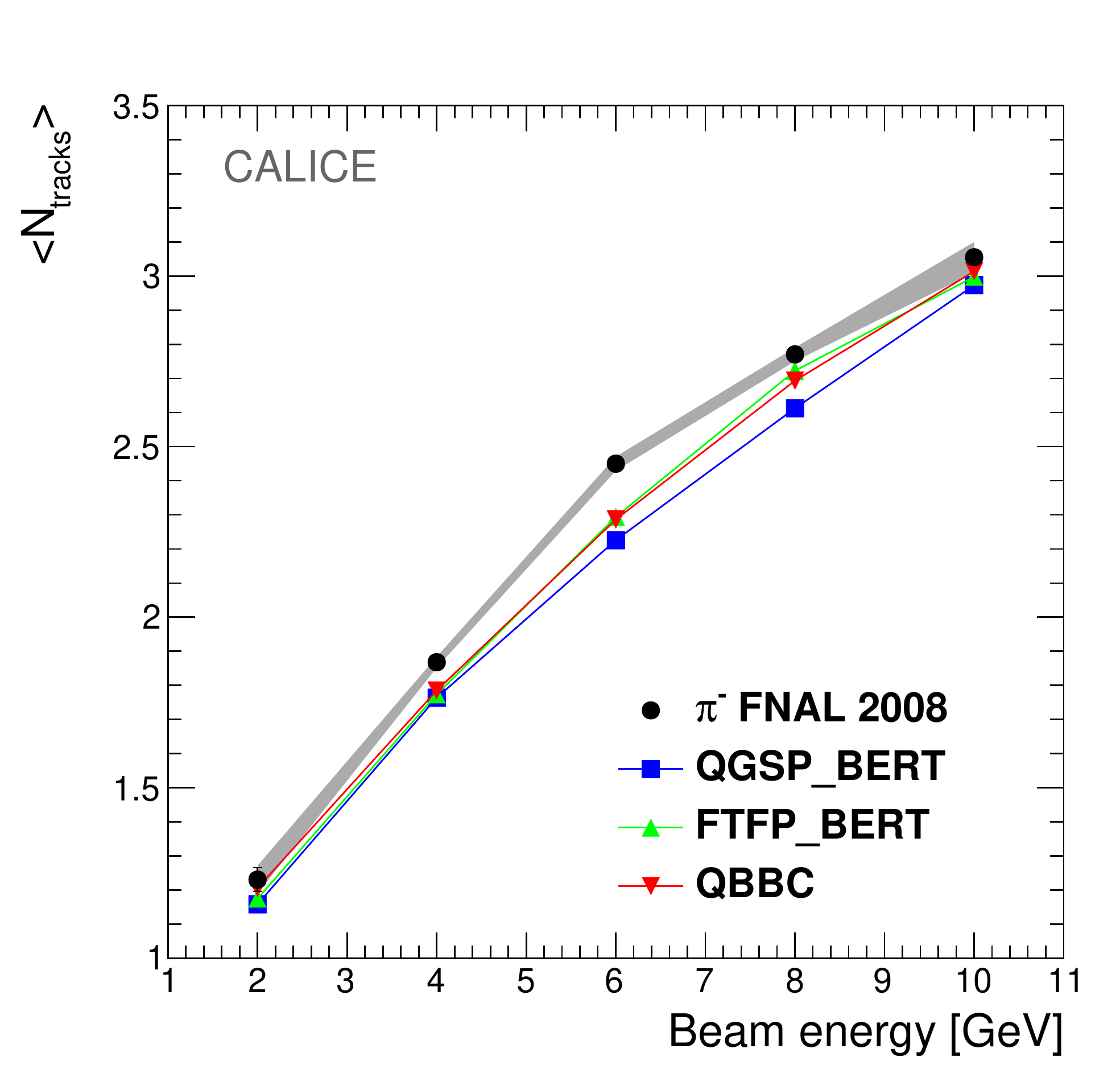}
\caption{\label{fig:fulltrackgraph} \sl The average number of secondary tracks $\left< N_{\mathrm{tracks}} \right>$ as a function of the beam energy. Other details follow those of Fig.~\ref{fig:irgraph}. 
%for data (black points with error shaded band) in comparison to the three simulation models, \qgsp\ (blue squares), \ftfp\ (green upward-pointing triangles) and \qbbc\ (red downward-pointing triangles).  Error bars represent statistical errors and the error band the systematic error from the correction for double $\uppi^-$ background.
}
\end{figure}

Figure \ref{fig:fulltrackgraph} shows the dependence of $\left<N_{\mathrm{tracks}}\right>$ on the beam energy for data and the simulation models. With increasing beam energy the centre-of-mass energy available for the $\uppi$-tungsten scattering increases with the square-root of the beam energy according to fixed target kinematics. It is therefore expected that the number of outgoing tracks increases correspondingly. This is indeed observed in data and simulation. The approximately linear increase at smallest energies flattens out towards higher beam energies. The extension of the interaction zone also increases with energy, see for example Sec.~\ref{sec:latrad}. This makes it more and more difficult to reconstruct clean tracks in the finite volume of the \ecal. The simulation models are in agreement with the data at beam energies of 2\,GeV and 10\,GeV and underestimate the number of secondary tracks by up to 7\% at intervening energies. 
%For convenience Fig.~\ref{fig:trackgraph96} shows the comparison between data and two physics lists as contained in the \geantfour\ Version 9.6, see Ref.~\cite{bib:can-055}. In the intermediate energy range, i.e. between 4 and 8\,GeV, these physics list are slightly further away from the data in Version 10.1 than this was the case in Version 9.6.

%Both simulation samples are in agreement with data within systematic uncertainties here given by the variation of the \ep\ according to $\varepsilon = 0.03 \pm 0.01$.

%As a measure for the sensitivity of the reconstructed number of tracks on the actual value of the \ep, the estimator 
%\begin{equation}
%\Delta {\cal O} = < {\cal O}(\varepsilon_{up}) - {\cal O}(\varepsilon_{low}> /{\cal O}(\varepsilon_{nom} = 0.03)>
%\end{equation}
%is introduced
%The sensitivity to the \ep\ defined by Eq.~\ref{eq:sens} for ${\cal O}=N_{tracks}$, $\varepsilon_{1},=0.04$, $\varepsilon_{2},=0.02$ and $\varepsilon_{nom.}=0.03$   
%$\Delta N_{tracks} = <N_{tracks}(\varepsilon = 0.04) - N_{tracks}(\varepsilon = 0.02)> /<N_{tracks}(\varepsilon = 0.03)> $ is used. 
%is shown in Fig. \ref{fig:dtracksgraph}. Within the chosen range the number of reconstructed tracks varies by about 10\% for both, data and the three {\sc Geant4} physics lists. 

%\begin{figure}[H]
%\centering
%\begin{subfigure}{0.5\textwidth}
%\centering
%\includegraphics[width=.90\linewidth]{stdselection/ntracks-graph.pdf}
%\caption{\label{fig:trackgraph10} }
%\end{subfigure}% 
%\begin{subfigure}{0.5\textwidth}
%\centering
%\includegraphics[width=.90\linewidth]{stdselection/ntracks-graph-v96.pdf}
%\caption{\label{fig:trackgraph96} }
%\end{subfigure}
%\caption{\label{fig:fullitrackgraph} \sl  
%The average number of secondary tracks $\left< N_{tracks} \right>$ as as function of the beam energy for data (black points with error shaded band) in comparison to the three physics lists, \qgsp\ ( blue squares), \ftfp\ (green upward-pointing triangles) and \qbbc\ (red downward-pointing triangles) generated with version 10.1 (a) and in comparison to the \qgsp\ and \ftfp\ models generated with version 9.6 (b).  Error bars represent statistical errors and the error band the systematic error from the correction for double $\uppi^-$ events.}
%\end{figure}

%|||||||||||||||||||||Hit distribution||||||||||||||||||||||

\subsection{Number of hits per track}
\begin{figure}[H]
\centering
\begin{subfigure}{0.5\textwidth}
\centering
\includegraphics[width=.90\linewidth]{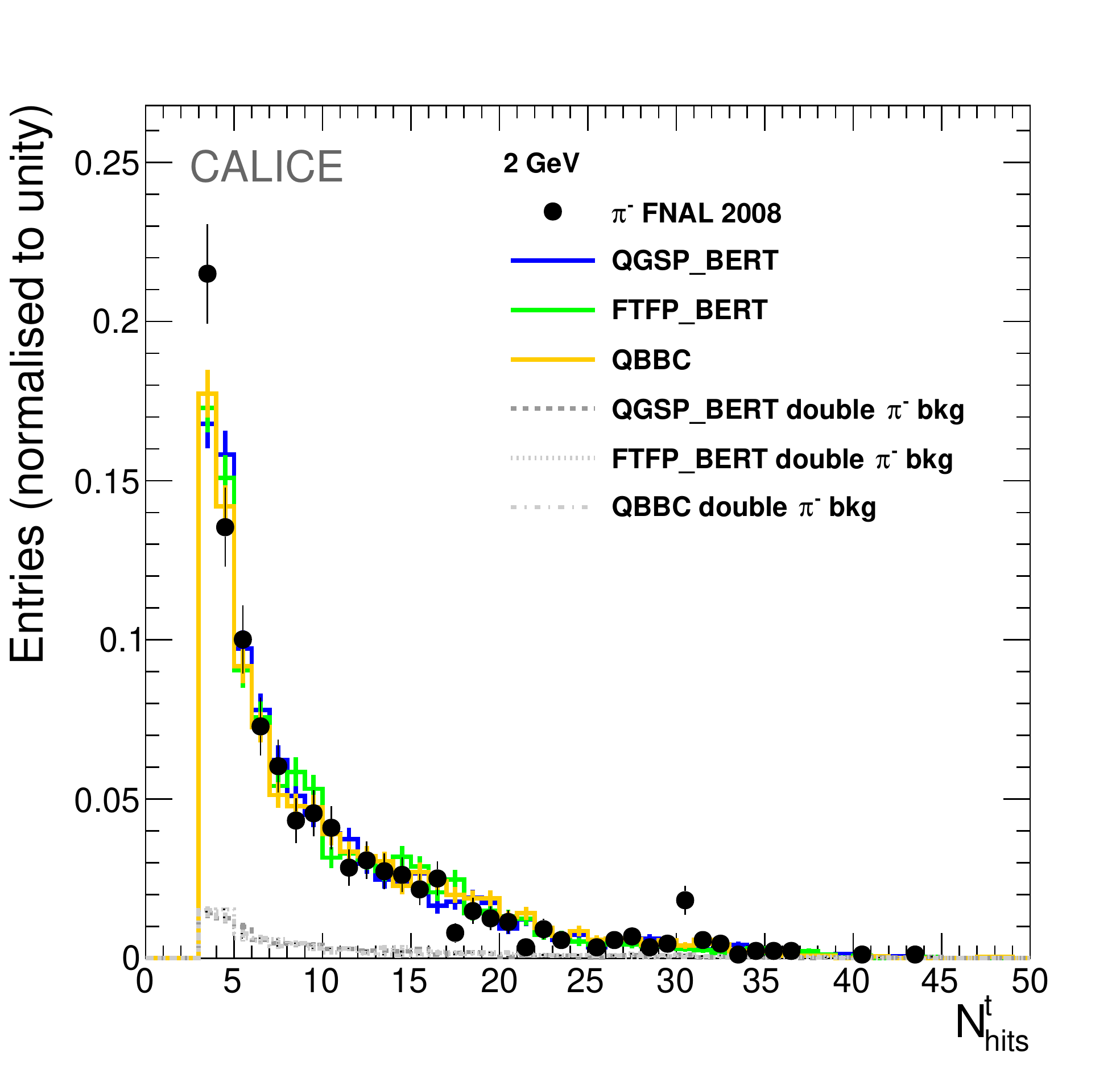}
\caption{\label{fig:trh2} }
\end{subfigure}% 
\begin{subfigure}{0.5\textwidth}
\centering
\includegraphics[width=.90\linewidth]{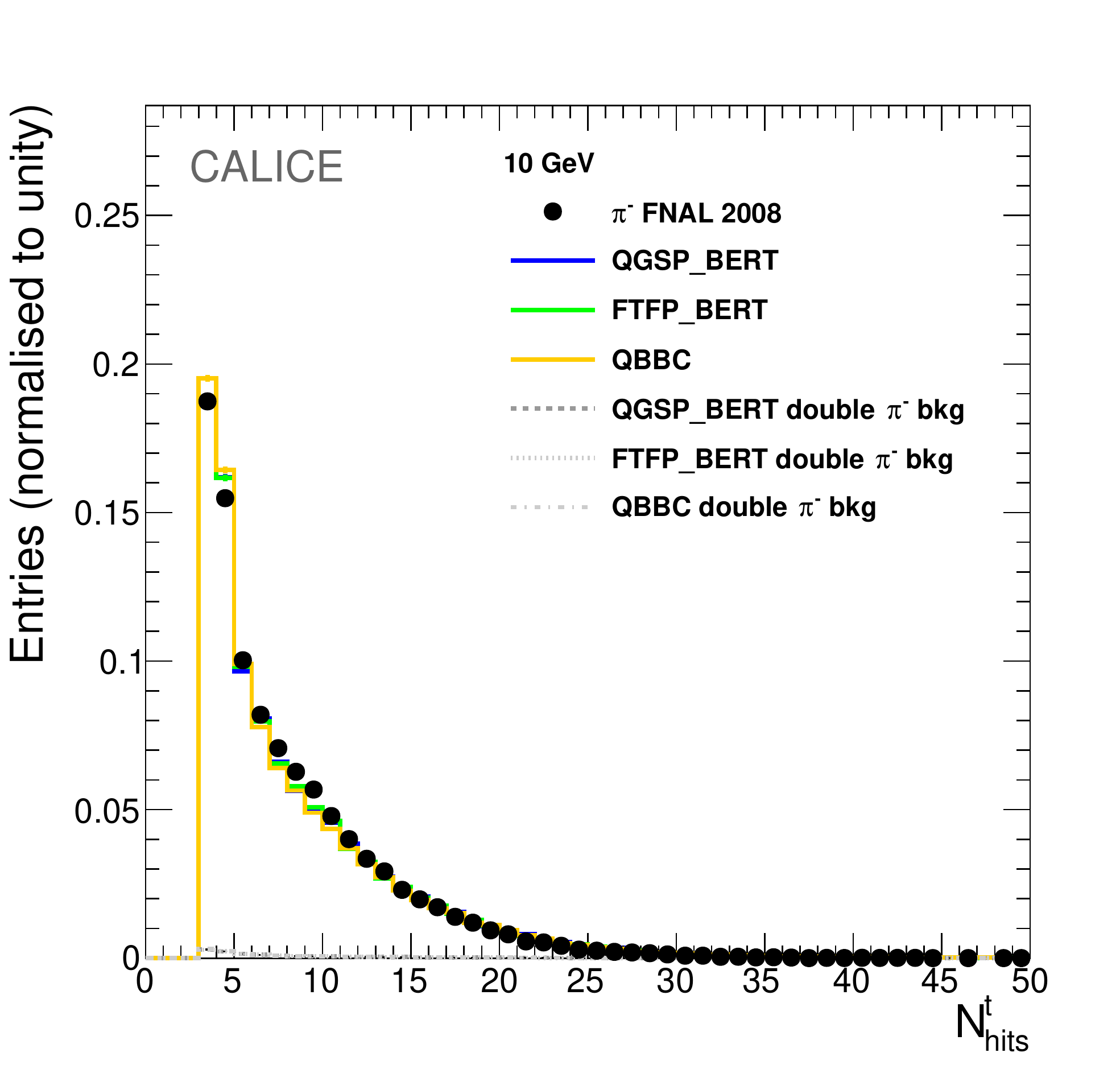}
\caption{\label{fig:trh10} }
\end{subfigure}
\caption{\label{fig:trackhitsexample} \sl Number of hits per reconstructed track for energies of 2 GeV (a) and 10 GeV (b) of the beam energy. Other details follow those of Fig.~\ref{fig:clusterexample}.   
%as observed in data (points with error bars) and for the three simulation models, \qgsp\ (blue histogram), \ftfp\ (green histogram) and \qbbc\ (yellow histogram). The double $\uppi^-$ background for the three models is shown by the grey dashed, dotted and dash-dotted histograms. All histograms are normalised to unity. Error bars represent statistical uncertainties only.
}
\end{figure}

The number of hits per track $N_{\mathrm{hits}}^\mathrm{t}$ is an essential characteristic of the reconstructed tracks. 
The histograms of $N_{\mathrm{hits}}^\mathrm{t}$ for 2 and 10\,GeV beam energy are shown in Fig. \ref{fig:trackhitsexample}. The distributions obtained for data and Monte Carlo are in good agreement with each other. 

\begin{figure}[H]
\centering
\includegraphics[width=.50\linewidth]{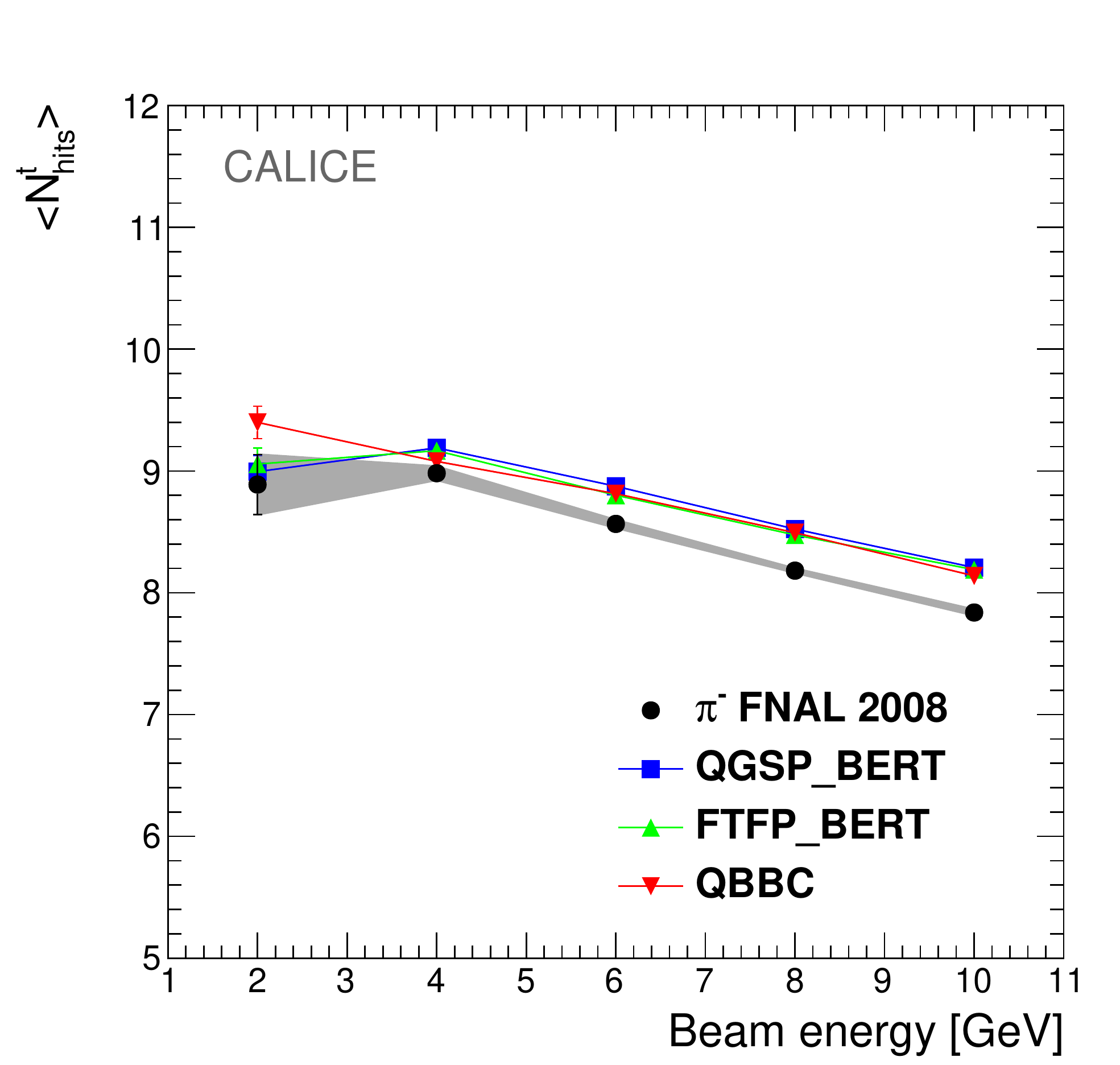}
\caption{\label{fig:fulltrackhitsgraph} \sl The average number of hits per reconstructed track  $\left<N_{\mathrm{hits}}^t\right>$ as a function of the beam energy. Other details follow those of Fig.~\ref{fig:irgraph}. 
%\caption{\label{fig:fullitrackgraph} \sl  
%for data (black points with error shaded band) in comparison to the three simulated models, \qgsp\ (blue squares), \ftfp\ (green upward-pointing triangles) and \qbbc\ (red downward-pointing triangles).  Error bars represent statistical errors and the error band the systematic error from the correction for double $\uppi^-$ events.
}
\end{figure}

Figure \ref{fig:fulltrackhitsgraph} shows the dependence of $\left<N_{\mathrm{hits}}^\mathrm{t}\right>$ on the beam energy for data and the simulation models. Data and simulation agree within 5\%.  For energies greater than 4\,GeV all simulation models are, however, systematically above the data. Note that the average number of hits slightly decreases with increasing energy. The increasing size of the interaction zone limits the space available for track reconstruction. This observation is, therefore, consistent with the flattening of the number of tracks observed in Sec.~\ref{sec:ntracks}.  
%The parametric uncertainty for the number of hits per track is defined as $\Delta N_{hits}^t = <N_{hits}^t(\varepsilon = 0.04) - N_{hits}^t(\varepsilon = 0.02)> /  <N_{hits}^t(\varepsilon = 0.03)>$. 
%The sensitivity to the \ep\  as defined by Eq.~\ref{eq:sens} for ${\cal O}=N_{hits}$, $\varepsilon_{1},=0.04$, $\varepsilon_{2},=0.02$ and $\varepsilon_{nom.}=0.03$ is shown in Fig. \ref{fig:dtrackshitsgraph}. For the chosen parameter range the sensitivity increases with increasing beam energy from 1\% to about 5\% for both, data and the three {\sc Geant4} physics lists. 
%Figure \ref{fig:dtrackshitsgraph} demonstrates a similar behavior of parametric uncertainty on $N_{hits}^t$ for data and Monte Carlo. 

%|||||||||||||||||||Angular distribution||||||||||||||||||||
\subsection{Angular distributions}
Due to the high granularity of the \ecal\, further tracking observables such as the polar ($\theta$) and azimuthal ($\phi$) angles of secondary tracks become available.
%In the absence of a magnetic field, the secondary particles from hadronic interaction undergo only multiple elastic scattering in the detector material. Therefore, the direction of the initial momentum coincides approximately with the direction of the track that is visible in the \ecal.
Both angles are measured in the right-handed coordinate frame defined in Sec.~\ref{sec:fnal} with $\theta$ measured relative to the $z$-axis. The track direction is calculated from the position of the first and the last hit of the track along the $z$-axis. 

\begin{figure}[H]
\centering
\begin{subfigure}{0.5\textwidth}
\centering
\includegraphics[width=.90\linewidth]{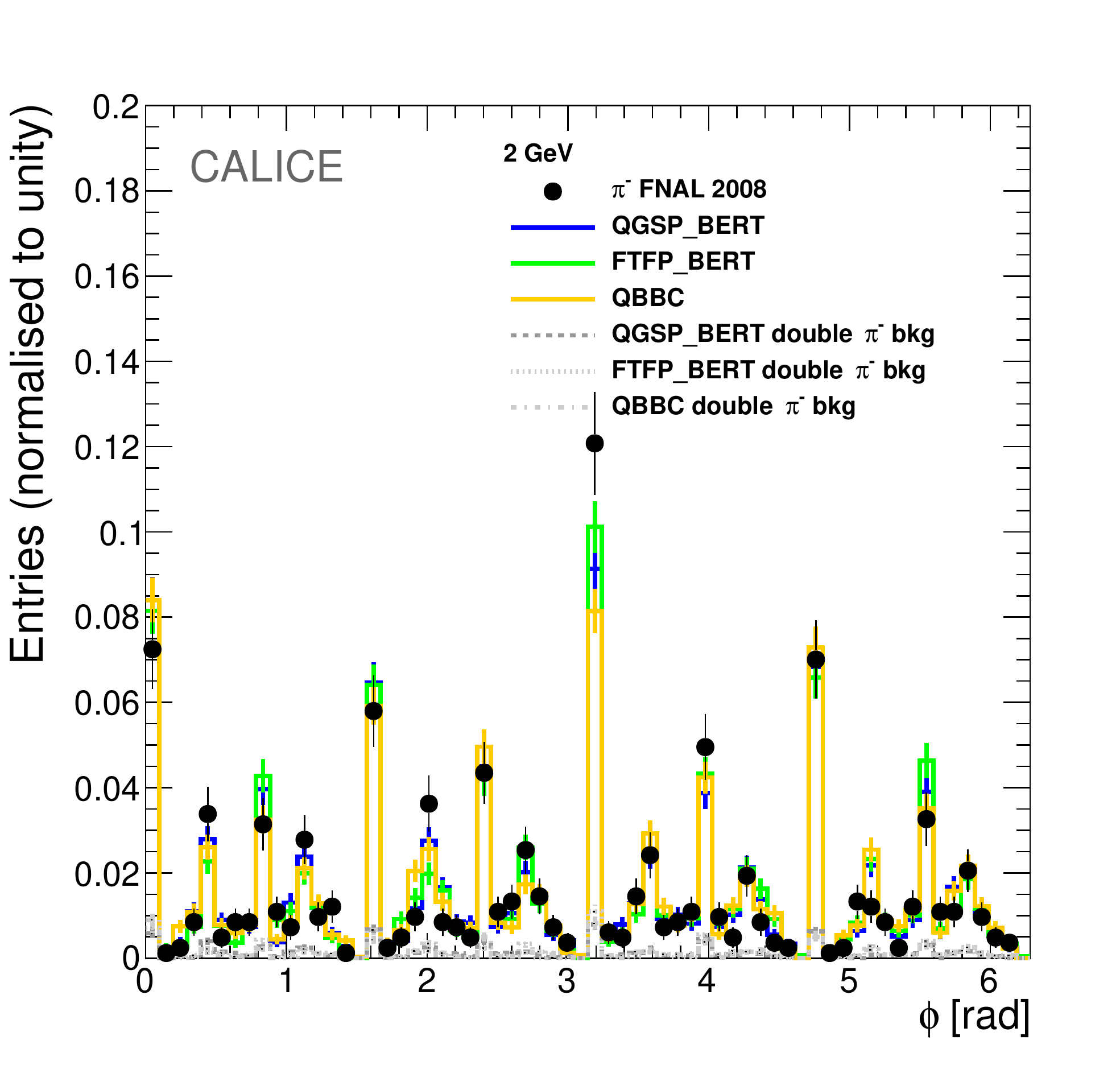}
\caption{\label{fig:phi2} }
\end{subfigure}% 
\begin{subfigure}{0.5\textwidth}
\centering
\includegraphics[width=.90\linewidth]{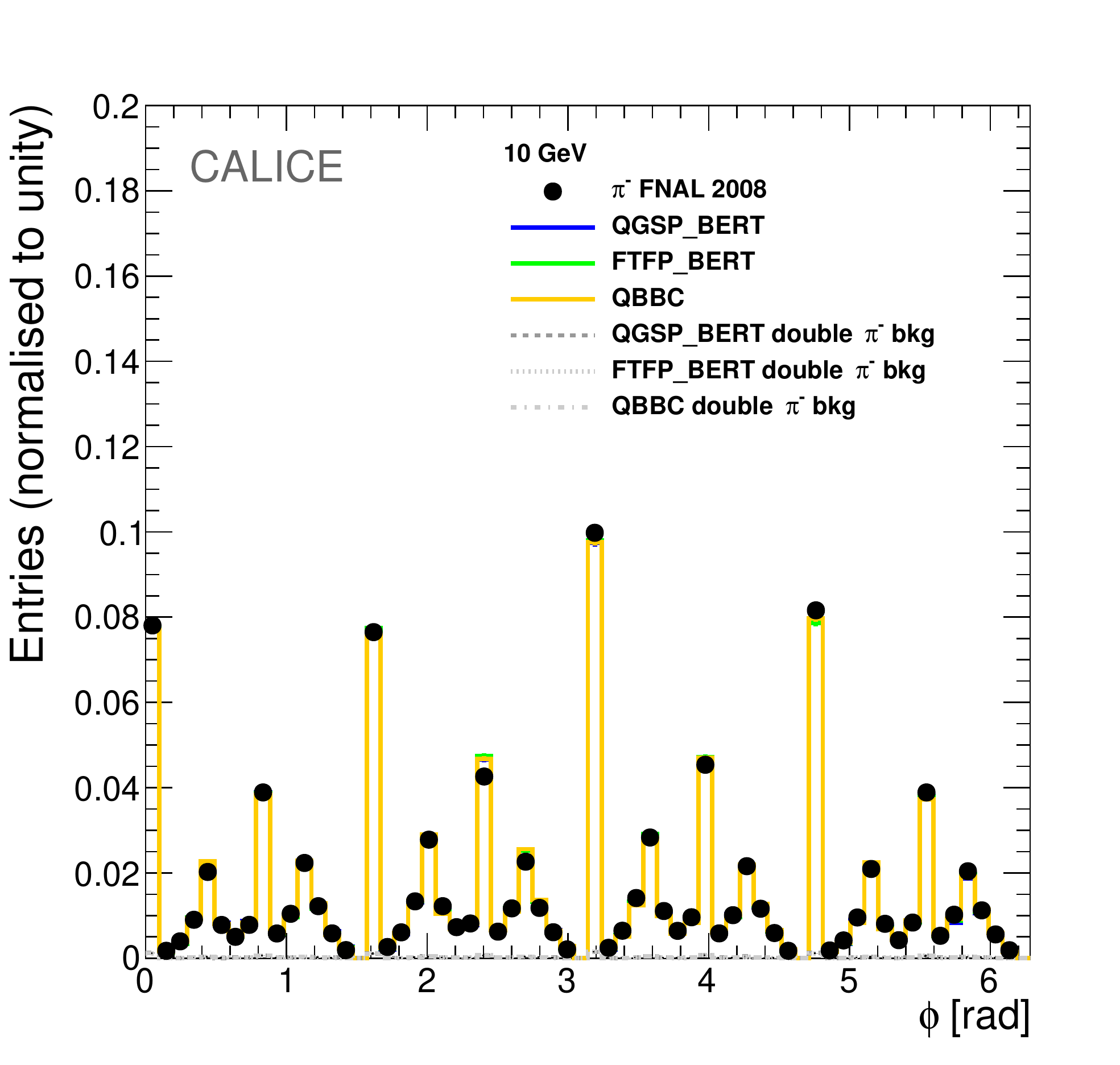}
\caption{\label{fig:phi10} }
\end{subfigure}
\caption{\label{fig:phiexample} \sl 
Comparison of the azimuthal angle $\phi$ of secondary tracks for energies of 2 GeV (a) and 10 GeV (b) of the incoming $\uppi^-$-mesons. Other details follow those of Fig.~\ref{fig:clusterexample}. 
%as observed in data (points with error bars) and for the three {\sc Geant4} physics lists, \qgsp\ (blue histogram), \ftfp\ (green histogram) and \qbbc\ (yellow histogram).  The double $\uppi^-$ background for the\ref{fig:phiexample} three models is shown by the grey dashed, dotted and dash-dotted histograms. All histograms are normalised to unity. Error bars represent statistical uncertainties only.
}
\end{figure}

\begin{figure}[H]
\centering
\begin{subfigure}{0.5\textwidth}
\centering
\includegraphics[width=.90\linewidth]{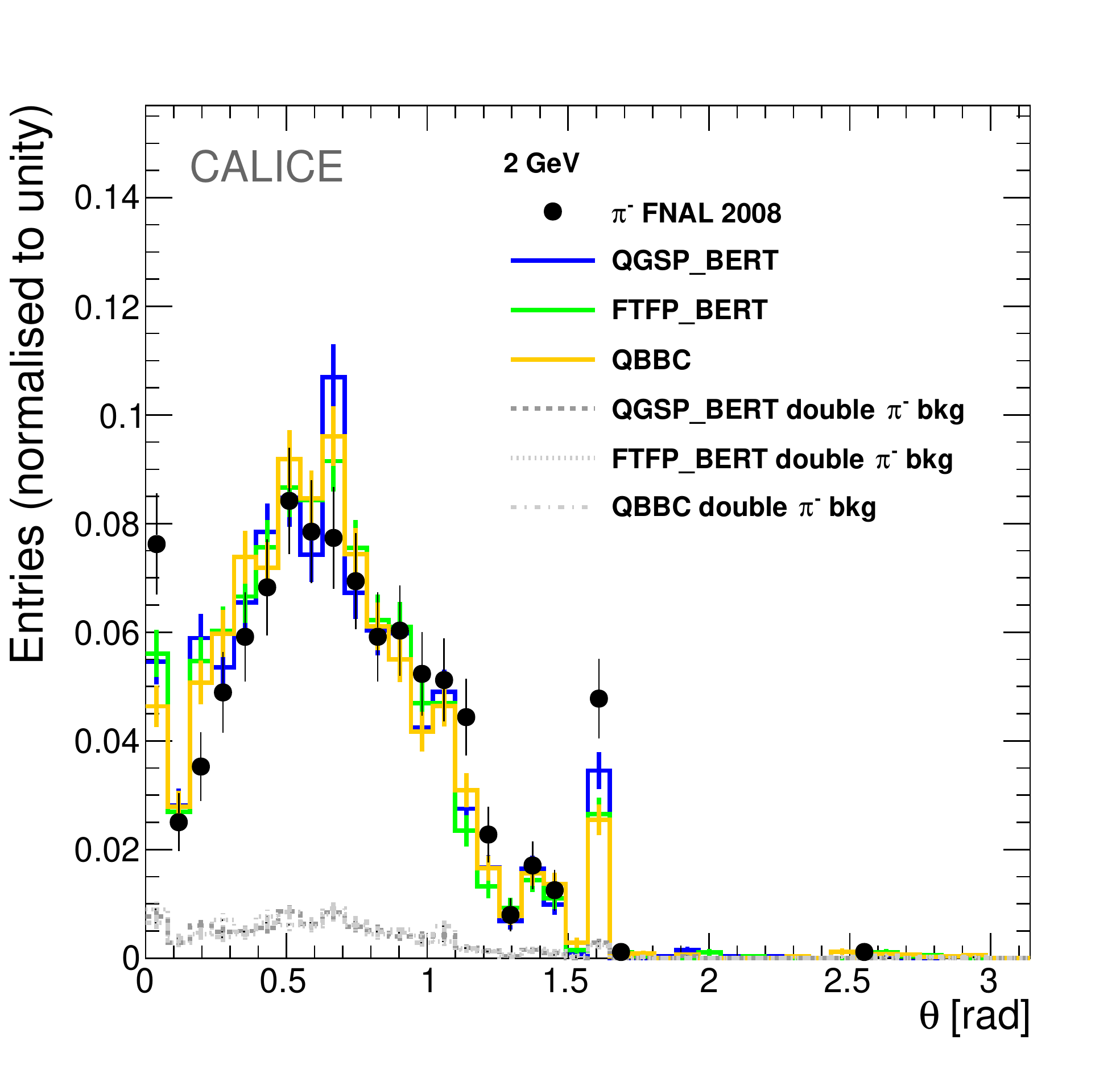}
\caption{\label{fig:theta2} }
\end{subfigure}% 
\begin{subfigure}{0.5\textwidth}
\centering
\includegraphics[width=.90\linewidth]{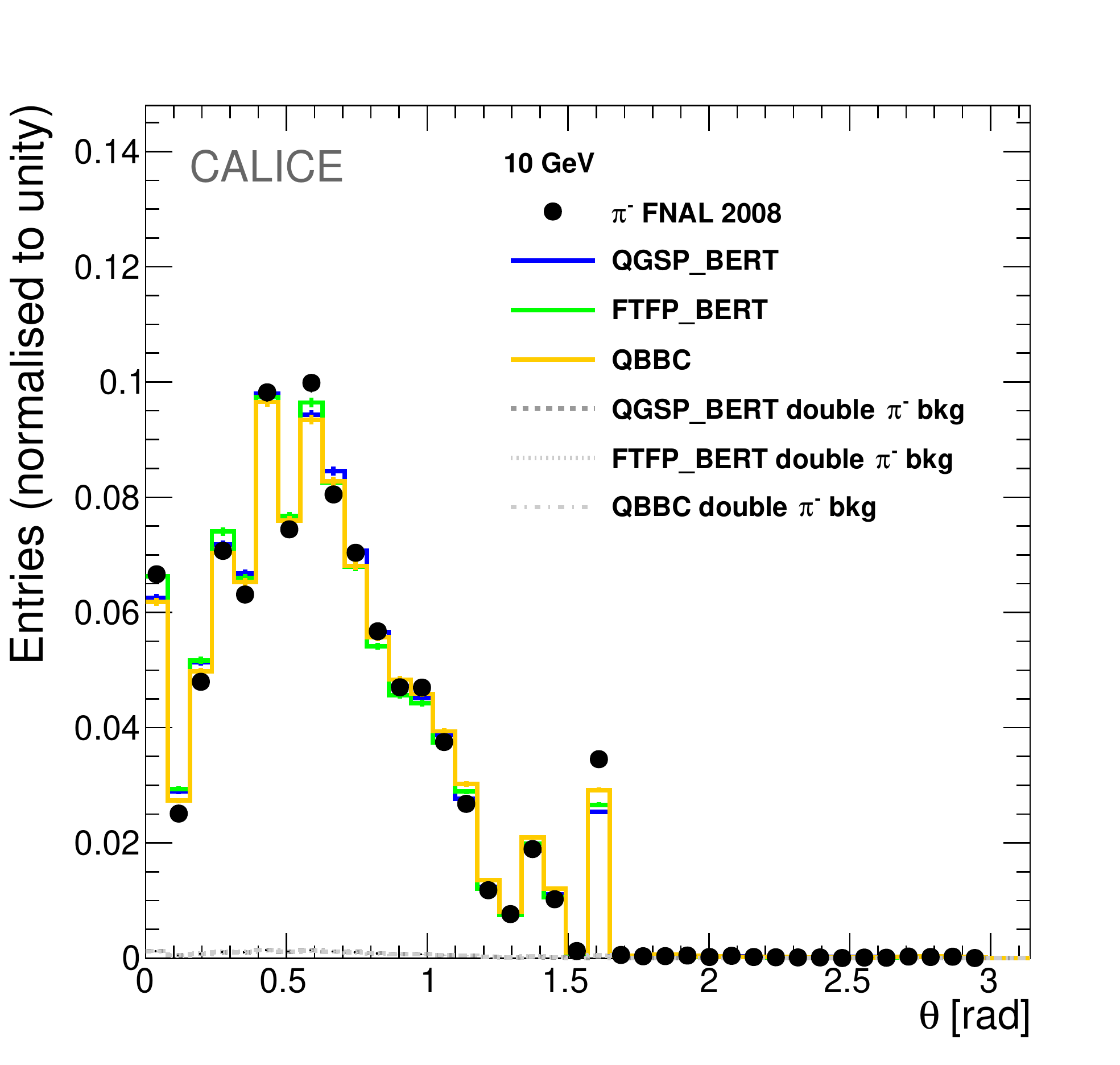}
\caption{\label{fig:theta10} }
\end{subfigure}
\caption{\label{fig:thetaexample} \sl 
The polar angle $\theta$ of secondary tracks for energies of 2 GeV (a) and 10 GeV (b) of the incoming $\uppi^-$-mesons. Other details follow those of Fig.~\ref{fig:clusterexample}. 
%as observed in data (points with error bars) and for the three {\sc Geant4} physics lists, \qgsp\ (blue histogram), \ftfp\ (green histogram) and \qbbc\ (yellow histogram)  The double $\uppi^-$ background for the three models is shown by the grey dashed, dotted and dash-dotted histograms. All histograms are normalised to unity. Error bars represent statistical uncertainties only.
}
\end{figure}

Figures \ref{fig:phiexample} and \ref{fig:thetaexample} display histograms of the $\phi$ and  $\theta$ angles, respectively, for 2 and 10\,GeV data together with corresponding corresponding results from simulation models. When corrected for the staggering of the detector layers in $x$~\cite{Anduze:2008hq}, the pad coordinates of the \ecal\ define a grid with a step width of about 1\,cm in the lateral direction. 
%The \tfa\ produces tracks with start and end points that are fixed to discrete calorimeter hit coordinates. 
This leads to a discretisation of the measured track direction. In particular, in the case of the azimuthal angle $\phi$, values that are a multiple of $\mathrm{\uppi/4}$ are favoured. Beyond that the distribution in $\phi$ is isotropic as expected. The bulk of the tracks are scattered in polar angles $\theta$ less than  $\mathrm{\uppi/2}$ as expected for a fixed target scattering. On the other hand the polar angle spectrum develops a long tail created by backward scattered particles. 
The simulation models reproduce the data adequately, largely within the experimental uncertainties.  
%For 2 and 10\,GeV beam energy the simulation models produce tracks with a similar angular distribution and reproduce the measured distributions adequately, most of the time within the experimental errors. 
%This gives evidence that the \ecal\ geometry is correctly implemented into the Monte Carlo simulation.

\begin{figure}[H]
\centering
\includegraphics[width=.50\linewidth]{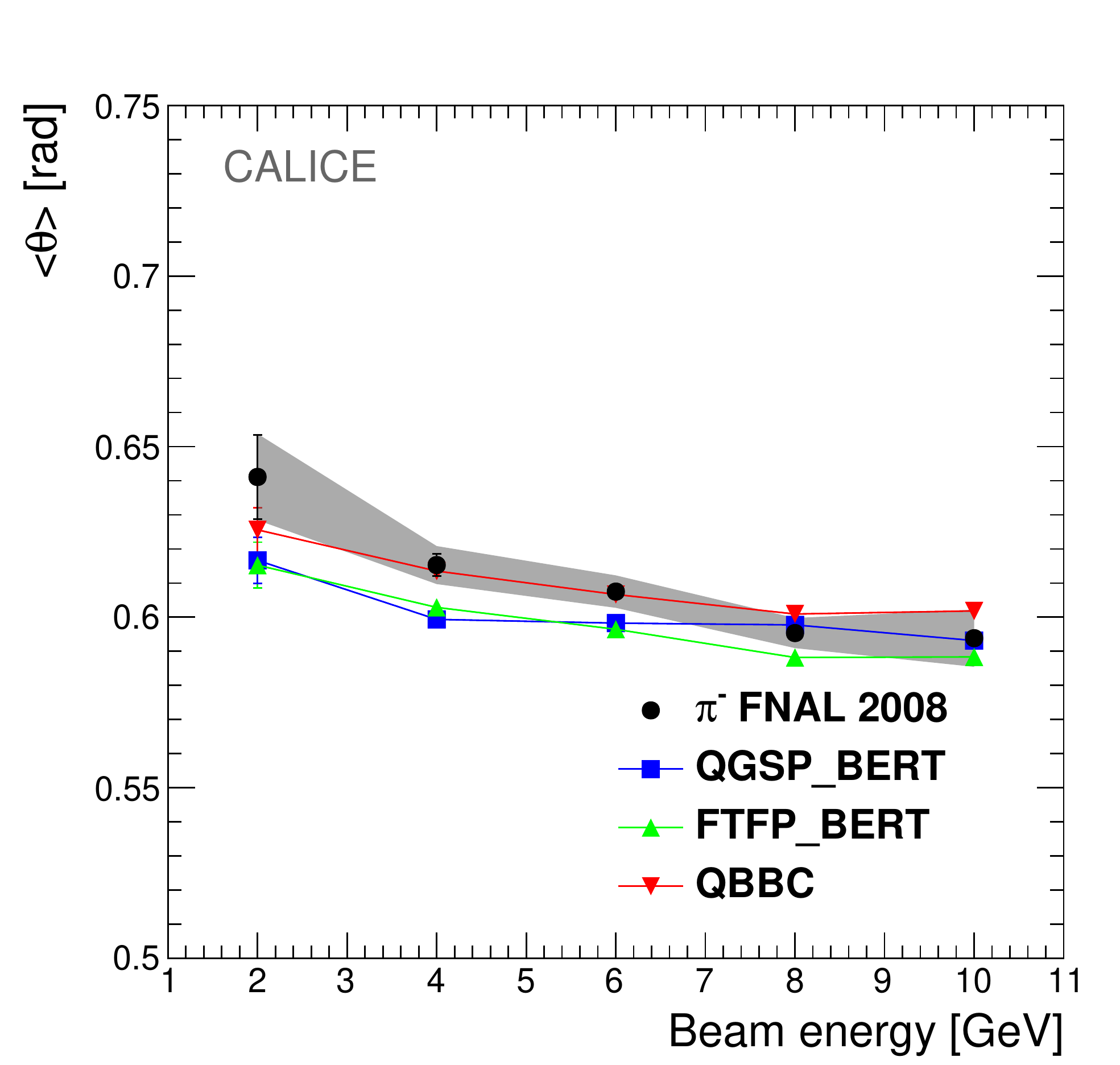}
\caption{\label{fig:fullthetagraph} \sl 
The truncated mean polar angle $\left<\theta\right>$ of secondary tracks as a function of the beam energy. Only tracks with polar angles less than $\mathrm{\uppi/2}$ have been selected. Other details follow those of Fig.~\ref{fig:irgraph}.  
%for data (black points with error shaded band) in comparison to the three simulated models, \qgsp\ (blue squares), \ftfp\ (green upward-pointing triangles) and \qbbc\ (red downward-pointing triangles).  Error bars represent statistical errors and the error band the systematic error from the correction for double $\uppi^-$ background.
}
\end{figure}

The truncated mean of the $\theta$ angle, $\left<\theta\right>$, which can be interpreted as a measure of the collimation of the secondary particles, is shown in Fig. \ref{fig:fullthetagraph} as a function of the beam energy. Here tracks with polar angles smaller than $\mathrm{\uppi/2}$ have been selected. 
%in order to be insensitive to fluctuations in the tail towards large polar angles, see Fig. \ref{fig:thetaexample}.  
The observable $\left<\theta\right>$ has only a weak dependence on the beam energy but shows the tendency to decrease with increasing energy as expected due to the increase of the boost transferred to the secondary particles.  The simulation models reproduce the data within a few percent, albeit the boost is less visible.  

%\begin{figure}[H]
%\centering

%\begin{figure}[H]
%\centering
%\begin{subfigure}{0.5\textwidth}
%\centering
%\includegraphics[width=.90\linewidth]{stdselection/theta-graph.pdf}
%\caption{\label{fig:thetagraph10} }
%\end{subfigure}% 
%\begin{subfigure}{0.5\textwidth}
%\centering
%\includegraphics[width=.90\linewidth]{stdselection/theta-graph-v96.pdf}
%\caption{\label{fig:thetagraph96} }
%\end{subfigure}
%\caption{\label{fig:fullthetagraph} \sl  
%The average polar angle $\left<\theta\right>$ of secondary tracé ks $\left< N_{tracks} \right>$ as as function of the beam energy for data (black points with error shaded band) in comparison to the three simulated models, \qgsp\ (blue squares), \ftfp\ (green upward-pointing triangles) and \qbbc\ (red downward-pointing triangles) generated with version 10.1 (a) and in comparison to the \qgsp\ and \ftfp\ models generated with version 9.6 (b).  Error bars represent statistical errors and the error band the systematic error from the correction for double $\uppi^-$ events.}
%\end{figure}

%A further discussion on the relationship between the \ep ,  the polar angle $\theta$ and the track length $l$ can be found in Appendix~\ref{app:b}.

%||||||||||||||||||||MPV energy deposition|||||||||||||||||||||
\subsection{Energy deposition by secondary tracks}

%The developed \tfa\ can select tracks from MIP-like particles in the \ecal\ that have no electron contamination.
%The secondary tracks in hadronic cascades should have the same mean energy deposition for any initial particle energy.
At energies relevant for this study, the secondaries that create sizeable tracks cross the detector volume behaving in a similar manner to minimally ionising particles. 
 %This motivates to investigate whether the secondary tracks can be used for an in-situ MIP calibration of the \ecal. 
This fact may be exploited as an in situ calibration of the detector, or at least to monitor the response of individual detector regions. For this specific study the selection criteria of events and tracks are modified as follows: 
\begin{itemize}
\item events are required to have more than one track and an interaction region to suppress soft inelastic scattering interactions at lower energies;
\item reconstructed tracks must have a length $l\geq 8$\,g.\,u.\,and $l / N_{\mathrm{hits}} > 0.9$\,g.\,u.\,to select long `pencil-like' tracks;
\item Reconstructed tracks must have a polar angle $\theta < 0.3$\,rad to reduce the angular dependence of the energy deposits.
\end{itemize}

\begin{figure}[H]
\centering
\begin{subfigure}{0.5\textwidth}
\centering
\includegraphics[width=.90\linewidth]{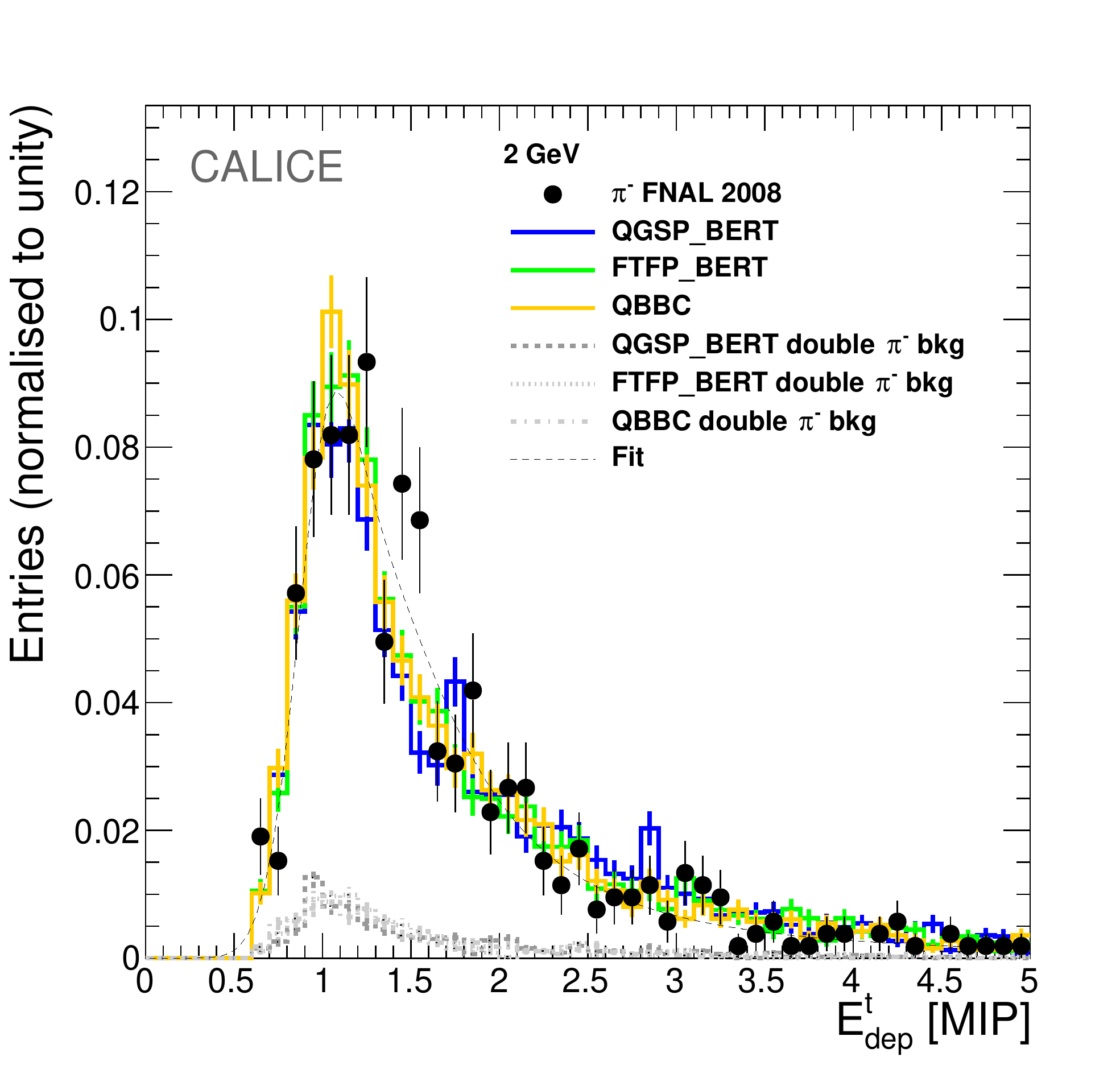}
\caption{\label{fig:calib2} }
\end{subfigure}% 
\begin{subfigure}{0.5\textwidth}
\centering
\includegraphics[width=.90\linewidth]{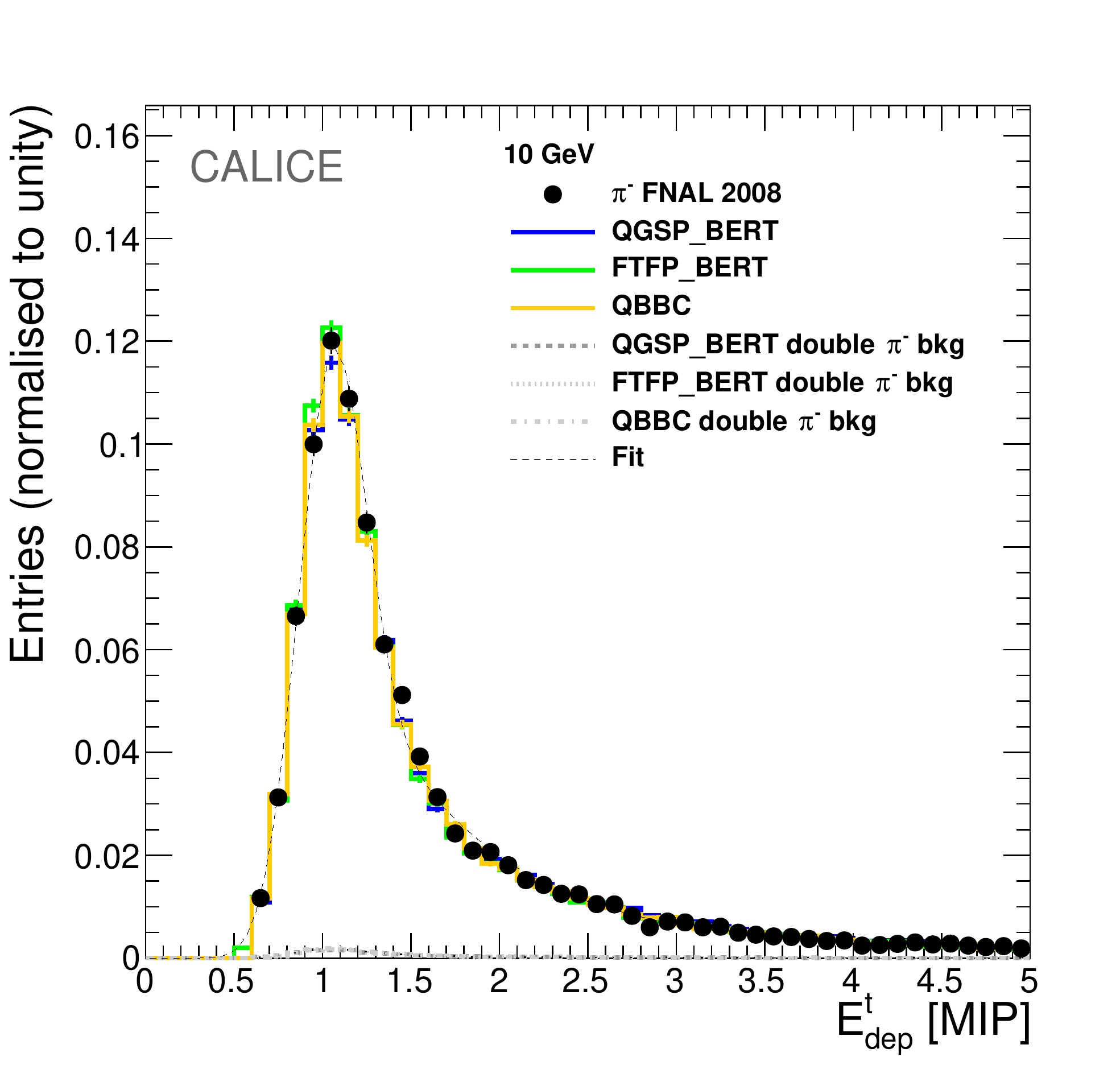}
\caption{\label{fig:calib10} }
\end{subfigure}
\caption{\label{fig:calib} \sl 
Energy deposition by secondary tracks observed in data (points with error bars) and for the three simulation models for beam energies of 2 GeV (a) and 10 GeV (b). The spectra are fit by the convolution of a Landau distribution with a Gaussian resolution function. The double $\uppi^-$-meson background for the three models is shown by the grey dashed, dotted and dash-dotted histograms, respectively. All histograms are normalised to unity. Error bars represent statistical uncertainties only.
}
\end{figure}

Figure~\ref{fig:calib} displays the energy deposition per hit by secondary tracks $E_{\mathrm{dep}}^\mathrm{t}$ for beam energies of 2 and 10\,GeV. Both distributions peak at around 1\,MIP as expected for straight MIP-like tracks. The overlaid fit is a Landau distribution convolved with a Gaussian resolution function, describing the data well. The tighter selection criteria reduce considerably the event sample at 2\,GeV. As a consequence, the statistical uncertainty of the fit is large for the 2\,GeV sample. %The data at 2\,GeV are thus discarded in the following.

\begin{figure}[H]
\centering
\includegraphics[width=0.5\textwidth]{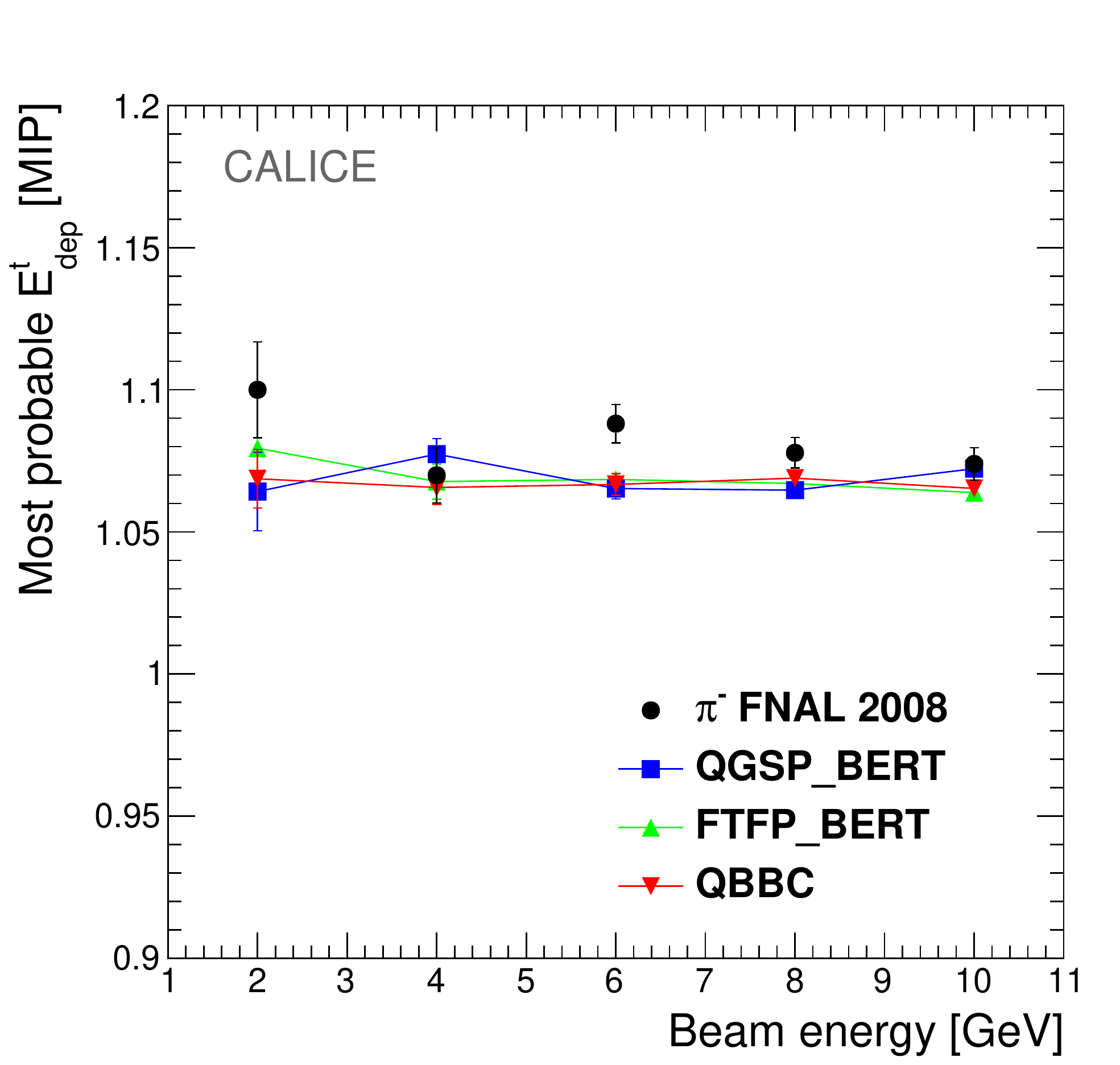}
\caption{\label{fig:calibrationgraph} \sl MPV of the Landau fit to the $E_{\mathrm{dep}}^t$ distributions of the `pencil-like` secondary tracks as a function of the beam energy for $\uppi^-$ data (black points with error bars) in comparison to the three simulation models. Error bars represent the statistical fit uncertainty.}
\end{figure}

Figure \ref{fig:calibrationgraph} presents the dependence of the most probable value (MPV) of the energy deposition in secondary tracks on beam energy.  The MPV is about 1.05, which is compatible with the fact that the selected tracks cross the detector pads at a small angle.
It can be seen that the detector response is both uniform within 1--2\% for the analysed energy range in data and reproduced by the simulation models within 1--2\%.  

%This is a non-trivial result. It shows that the algorithm has indeed selected MIP-like secondary tracks, since the MIP scale is expected to be independent of the underlying simulation model and the detector response should be largely independent of the energy of the primary particle. 

The algorithm has selected particles with approximately minimally ionising behaviour. The uniform response supports the idea that the secondary tracks can be used for the in situ monitoring of the calibration. 

%-------------------------------------------------------------
%------------------------SUMMARY------------------------------
%-------------------------------------------------------------
\section{Summary and outlook}

This study reveals the large potential of the CALICE \ecalp\ to obtain a detailed picture of the interactions of hadrons with matter. 
The article describes basic ideas and the application of a new simple \tfa\ for the \ecal. This algorithm allows for the reconstruction of tracks produced by secondary particles created in the interaction of hadrons with the absorber material, and hence to study the interaction region of hadron showers in the \ecal. The \tfa\ produces a new set of observables, based on reconstructed tracks of secondary particles and the interaction region of the hadronic cascades. 
%The results are stable w.r.t. small variations of the main parameter of the \tfa.

Data recorded in test beams at FNAL in 2008 using $\uppi^-$-mesons with energies between 2 and 10\,GeV are compared with predictions from the simulation models \qgsp, \ftfp\ and \qbbc\ as implemented in \geantfour\ Version 10.1. The agreement between data and simulation varies with beam energy and the chosen physics observable. In most cases data and simulation models agree within 10\% without revealing a clear preference for one of the chosen simulation models. 
%In this context it is worthwhile to remind that the interaction region is systematically 10\% wider than it is the case for the Monte Carlo simulation.

The largest discrepancy between data and the simulation models is observed for the deposited energy in and the radius of the interaction region. The measured energy deposition in the interaction region is up to 15\% higher than predicted by the Monte Carlo simulation.
The distributions of the number of secondary tracks and the number of hits per track for data are described well by the simulation models. The distribution of the polar angle of the reconstructed tracks in the simulation agrees with data within 3--4 \% and the distribution of azimuthal angles is reproduced well by the simulation models in spite of the non-trivial detector geometry.

%The new observables are sensitive to the different hadronic models implemented in the physics lists. 
%The mean polar angle of detected tracks as a function of the beam energy has a visible transition between Fritiof and Bertini cascades in the \ftfp\ physics list, as well as between Bertini and LEP models in the \qgsp\ physics list. The same effects can be seen for other observables. 

Future work should aim at transferring the insights about the interaction region and the secondaries emerging from it to the optimisation of Particle Flow Algorithms.

A tighter track selection leads to long tracks by particles that show approximately minimally ionising behaviour. The detector response determined using these tracks is stable to about 1--2\% over the tested energy range and shows good agreement with simulation. This observation can be exploited as a starting point for a feasibility study of an in situ calibration (or at least a regular monitoring of the detector) by means of the selected tracks.  

\section*{Acknowledgements}
%Tu peux faire ca avant la migration. 
%{\it [Data-set specific thanks to laboratories, technical staff, non-CALICE institutes providing equipment, ... --- not added now ---  ]}
We gratefully acknowledge the DESY, CERN and FNAL managements for their support and hospitality, and their accelerator staff for the reliable and efficient beam operation.
This work was supported 
by the FWO, Belgium; 
by the Natural Sciences and Engineering Research Council of Canada;
by the Ministry of Education, Youth and Sports of the Czech Republic;
%by the European Union's Horizon 2020 Research and Innovation programme under Grant Agreement 654168; %% AIDA-2020
%by the European Commission within Framework Programme 7 Capacities, Grant Agreement 262025;  %% AIDA (not 2020)
by the P2IO LabEx in the framework 'Investissements d'Avenir' managed by the French National Research Agency (ANR) under Grant Agreements ANR-10-LABX-0038 and ANR-11-IDEX-0003-01; 
by the 'Quarks and Leptons' Programme of CNRS/IN2P3 France;
by the Alexander von Humboldt Stiftung (AvH), Germany;
by the Bundesministerium f\"ur Bildung und Forschung (BMBF), Germany; 
by the Deutsche Forschungsgemeinschaft (DFG), Germany; 
by the Helmholtz-Gemeinschaft (HGF), Germany; 
by the I-CORE Program of the Planning and Budgeting Committee, Israel;
by the Nella and Leon Benoziyo Center for High Energy Physics, Israel;
by the Israeli Science Foundation, Israel;
by the National Research Foundation of Korea;
by the Korea-EU cooperation programme of National Research Foundation of Korea, Grant Agreement 2014K1A3A7A03075053; 
by the Netherlands Organisation for Scientific Research (NWO);
by the Science and Technology Facilities Council, UK;
by the Nuclear Physics, Particle Physics, Astrophysics and Cosmology Initiative, a Laboratory Directed Research and Development program at the Pacific Northwest National Laboratory, USA.
\section*{References}
\begin{footnotesize}
\bibliographystyle{utphys_mod}
\bibliography{had-pap}
%\bibliographystyle{hep}

%\renewcommand\refname{References}
%\end{thebibliography}
\end{footnotesize}

\end{document}